\documentclass{aastex631}
\pdfoutput=1 

\usepackage{comment}
\usepackage{amsmath}
\usepackage{threeparttable}

\submitjournal{ApJ}
\shorttitle{X-ray AGN-RPS connection}
\shortauthors{Tiwari et al.}

\begin{document}
\title{Can AGN activity be enhanced by ram pressure stripping? -- X-ray perspective}

\correspondingauthor{Juhi Tiwari}
\email{E-mail: jt0189@uah.edu}
\correspondingauthor{Ming Sun}
\email{E-mail: ms0071@uah.edu}

\author{Juhi Tiwari}
\affiliation{Department of Physics and Astronomy, The University of Alabama in Huntsville, 301 Sparkman Dr NW, Huntsville, AL 35899, USA}
\author{Ming Sun}
\affiliation{Department of Physics and Astronomy, The University of Alabama in Huntsville, 301 Sparkman Dr NW, Huntsville, AL 35899, USA}
\author{Rongxin Luo}
\affiliation{School of Physics and Electronic Science, Guizhou Normal University, Guiyang 550001, PR China}
\affiliation{Guizhou Provincial Key Laboratory of Radio Astronomy and Data Processing, Guizhou Normal University, Guiyang 550001, PR China}
\author{Matteo Fossati}
\affiliation{INAF-Osservatorio Astronomico di Brera, via Brera 28, I-20121 Milano, Italy}
\author{Chen, Chien-Ting J.}
\affiliation{Science and Technology Institute, Universities Space Research Association, Huntsville, AL 35805, USA; \\Astrophysics Office, NASA Marshall Space Flight Center, ST12, Huntsville, AL 35812, USA}
\author{Prathamesh Tamhane}
\affiliation{Department of Physics and Astronomy, The University of Alabama in Huntsville, 301 Sparkman Dr NW, Huntsville, AL 35899, USA}

\begin{abstract}
Ram pressure stripping (RPS) is an important process that plays a significant role in shaping the evolution of cluster galaxies and their surrounding environment. Despite its recognized significance, the potential connection between RPS and AGN activity in cluster galaxies remains poorly understood. Recent claims, based on optical emission line diagnostics, have suggested such a connection. Here, we investigate this relationship from an X-ray perspective using a sample of galaxies undergoing RPS in four nearby galaxy clusters -- A1656, A1367, A426, and A3627. This study is the first to test such a connection from an X-ray standpoint. Our analysis reveals no signs of enhanced X-ray AGN activity in our sample, with most RPS galaxies ($\sim 90\%$) showing X-ray luminosities below $10^{41}$ erg s$^{-1}$ in their central point sources. Moreover, there is no noticeable difference in X-ray AGN activity among RPS galaxies compared to a control sample of non-RPS galaxies, as demonstrated by similar X-ray luminosities observed in their central point sources. While the most luminous X-ray AGN in our sample is found in ESO 137-002, a galaxy undergoing RPS in A3627, there is no evidence for a widespread enhancement of X-ray AGN activity due to RPS. Given the limited sample size of our study, this could also indicate that either the X-ray AGN enhancement from RPS is at most weak, or the timescale for the X-ray AGN enhancement is short. This emphasizes the need for further investigations with larger X-ray samples to better understand the impact of RPS on AGN activity in cluster galaxies.
\end{abstract}


\section{Introduction} \label{sec:intro}
Active Galactic Nuclei (AGN) are some of the most energetically intense phenomena observed in the Universe. Typically situated in the centers or nuclei of massive galaxies (stellar mass $\gtrsim 10^{9.5}$ solar mass), they derive their power from the accretion of material onto a central supermassive black hole (SMBH) (see review by \citealt{kormendy2013} for general history and black hole detections). AGN rank among the most luminous objects in the Universe, emitting radiation across a broad spectrum -- from radio waves to X-rays and gamma rays (refer to \citealt{brandt2015} and \citealt{padovani2017} for a detailed overview of multiwavelength properties of AGN). Beyond their luminosity, AGN play a crucial role in galactic evolution, influencing the surrounding gas and impacting star formation in host galaxies through powerful outflows, known as feedback (e.g., \citealt{mcnamara2007, fabian2012, harrison2017}; also, see \citealt{cresci2015} for positive feedback in NGC 5643 and \citealt{shin2019} for both positive and negative feedback in NGC 5728). Additionally, studying AGN provides invaluable insights into the extreme environments surrounding SMBHs, allowing us to explore accretion processes and understand SMBH formation and growth (e.g., \citealt{miller2007, kormendy2013, heckman2014}).

The availability of cold gas to a central SMBH is a crucial factor in AGN activation and sustenance (e.g., \citealt{sabater2015, izumi2016, yesuf2020, fujita2023a, fujita2023b}). Various mechanisms can facilitate the funneling of gas towards the galactic centre, thereby triggering AGN. Commonly proposed mechanisms include galaxy-galaxy interactions and mergers where gravitational forces can disturb the distribution of gas within galaxies, causing it to lose angular momentum and fall towards the centre (e.g., \citealt{alonso2007, ellison2011, silverman2011, satyapal2014, goulding2018, gao2020}). In addition to these, the environment in which a galaxy resides can also have substantial impact on its AGN activity. For example, in the central galaxies of cool-core galaxy clusters, AGN activity can be potentially fueled by the presence of large amounts of radiatively cooling, dense gas at the cluster centres (e.g., \citealt{best2007, mcnamara2007, DV22}). Another environmental mechanism that possibly affects AGN activity in cluster galaxies is ram pressure stripping (RPS), a process in which galaxies moving through the ambient medium experience a drag force that strips away their interstellar gas. The removal of the interstellar gas from a galaxy due to RPS may reduce the gas reservoir available and decrease AGN activity (e.g., \citealt{delpino2023}). On the other hand, some recent observational (e.g., \citealt{poggianti2017_nature, peluso2022}) and simulation works (e.g., \citealt{marshall2018,ricarte2020}) indicate that AGN activity in galaxies can actually be enhanced by RPS. Other simulation works indicating that RPS could plausibly feed a central black hole include studies by \citet{farber2022} and \citet{akerman2023}. There are also studies which indicate no link between AGN activity and RPS (e.g., \citealt{roman-oliveira2019}; also see \citet{boselli2022} for a thorough overview of RPS in high-density environments). Currently, this controversy is still unresolved.

A valuable approach to investigate the influence of galaxy environment on AGN activity is to examine the AGN fraction -- the proportion of galaxies hosting an AGN within a specific population or environment. A significant body of literature suggests that the AGN fraction tends to be lower in galaxy clusters than in the field, based on both optical and X-ray studies. Early works, such as those by \citet{gisler1978} and \citet{dressler1985}, indicated that AGN selected using optical emission lines are less prevalent in clusters than in the field. Subsequent studies based on individual clusters as well as large surveys have confirmed these findings, revealing that the fraction of galaxies hosting an AGN, observed optically (bright or faint), is suppressed in regions with high local galaxy density (\citealt{kauffmann2004} ($0.03 < z < 0.1$); \citealt{coldwell2006} ($z <0.2$), \citealt {popesso2006} ($z < 0.08$), \citealt{gilmour2007} ($0.070 < z < 0.084$), \citealt{mahajan2010} ($0.015 < z< 0.03$), \citealt{sabater2013, sabater2015} ($0.03 < z <0.1$), \citealt{lopes2017} ($z\leq 0.1$)). We note that most of these studies investigate the optical AGN fraction across all emission-line galaxies (including early and late-types) combined. However, \citet{lopes2017} segregated their galaxy sample into early-type and late-type galaxies. They found that the AGN fraction decreases in denser cluster environments for both types of galaxies, although there is considerable scatter in the late-type population. The authors concluded that the observed variation in AGN activity with environment is not solely due to the morphology-density relation. Similar results emerge when AGN are detected based on X-ray signatures, indicating suppression of the X-ray luminous AGN fraction in rich cluster regions compared to the less rich galaxy groups and the field (\citealt{arnold2009} ($0.02 < z < 0.06$), \citealt{martini2009} ($z < 0.4$), \citealt{haines2012} ($0.15 < z < 0.30$), \citealt{koulouridis2010} ($0.073 < z < 0.279$), \citealt{ehlert2014} ($0.2 < z < 0.7$), \citealt{semyeong2014} ($0.5 < z < 1.1$), \citealt{koulouridis2018} ($0.1 < z < 0.5$), \citealt{mishra2020} ($z < 0.5$)). These findings support the scenario that AGN are mainly triggered by close galaxy interactions and mergers, which are more common in environments with low relative velocities, such as the field, low-mass groups, or cluster outskirts.

While optical observations are crucial for studying AGN, there are limitations associated with optical AGN selection. In certain galaxies, the central regions hosting the AGN may be obscured by gas and dust, posing a challenge for detecting the AGN in the optical spectrum. Additionally, in galaxies with high star formation rates or prominent stellar components, the light from the host galaxy can dominate, potentially overshadowing the AGN emission, especially in the optical wavelength range. These factors highlight the importance of complementary observations at other wavelengths, such as X-rays, to obtain a more comprehensive view of AGN activity and its contributing factors. AGN are prolific X-ray emitters, and point-like X-ray emission is one of the cleanest signatures of AGN, making X-ray observations particularly well-suited for AGN population studies (e.g, \citealt{ehlert2015}). Furthermore, X-ray selection of AGN is less affected by obscuration and less susceptible to orientation effects compared to optical selection, which may be strongly influenced by the viewing angle of the AGN, potentially introducing biases in the identified population.

There is also a longstanding debate on whether LINERs (or LIERs in some cases)\footnote{LINER: Low-Ionization Nuclear Emission-line Region;\\LIER: Low-Ionization Emission Region} should be classified as AGN (e.g., \citealt{yan2012}). While integral field spectroscopy is a powerful tool for obtaining spatially resolved spectra of galaxies and studying their nuclear regions, existing instruments (or studies) have not been able to probe the central $\lesssim$100 pc region of galaxies. For example, the MaNGA study by \citet{belfiore2016}, at a mean redshift of 0.03, examines the central $\sim$1 kpc region of galaxies despite the excellent spatial resolution of MaNGA. Belfiore’s study strongly supports the scenario in which LINERs are actually LIERs, with emission not originating from a central point source, but from diffuse stellar sources, likely hot, evolved (post-asymptotic giant branch) stars. Similarly, MUSE-IFU studies also probe the central $\sim$1 kpc regions of galaxies, limited by the seeing-dependent resolution of the instrument (e.g., \citealt{peluso2022} at median $z=0.05$). Therefore, while LIER emission might originate from nuclear sources such as low-luminosity AGN (in which case they would be considered LINERs), current instrumentation cannot resolve the central regions of galaxies down to 100 pc or smaller (for e.g., also see the  WiFeS IFU study of \citet{parkash2019}. As a result, other possible sources of LIER/LINER emission remain valid, leaving doubts about the true origin of these emissions. Additionally, it is worth noting that while IFU studies have gained traction in recent years, not many galaxies currently have IFU data available. High-resolution studies of the nuclear regions in a larger sample of galaxies are necessary to determine the true nature of LINER origin.

Given recent assertions suggesting that AGN activity in cluster galaxies is triggered by RPS (as discussed in paragraph 2 above), in this work, we investigate this proposition by examining the impact of RPS on AGN activity from an X-ray perspective. This is a timely project on a new controversy and there has never been a systematic X-ray study on this topic. In Section \ref{sec:galaxy sample, observations and data processing}, we detail the galaxy samples, observations, and data processing, followed by the presentation of results and discussion in Section \ref{sec:results and discussion}. Section \ref{sec:summary} provides a concise summary of our findings. Throughout this work, we adopt H$_{0}$=70 km s$^{-1}$ Mpc$^{-1}$, $\Omega$$_{M}$=0.3, and $\Omega$$_{\Lambda}$=0.7 for determining luminosity distances and magnitudes.

\section{Galaxy samples, observations and data processing}
\label{sec:galaxy sample, observations and data processing}
In this work, we aim to investigate whether there is an enhancement in X-ray AGN activity among cluster galaxies undergoing RPS. We begin our analysis with the study of X-ray AGN activity using \textit{Chandra} data in a sample of RPS galaxies from four nearby galaxy clusters, referred to as the `RPS sample'. Subsequently, we utilize data from the \textit{Sloan Digital Sky Survey} (SDSS) to build a `control sample' comprising galaxies not associated with galaxy clusters or groups but with intrinsic properties similar to those in the RPS sample. We then compare X-ray activity in galaxies from the two samples to elucidate the RPS-AGN connection. The description of our RPS and control galaxy samples, the observations used in this work, and the data processing methods employed are provided below.

\begin{figure}
\gridline{\fig{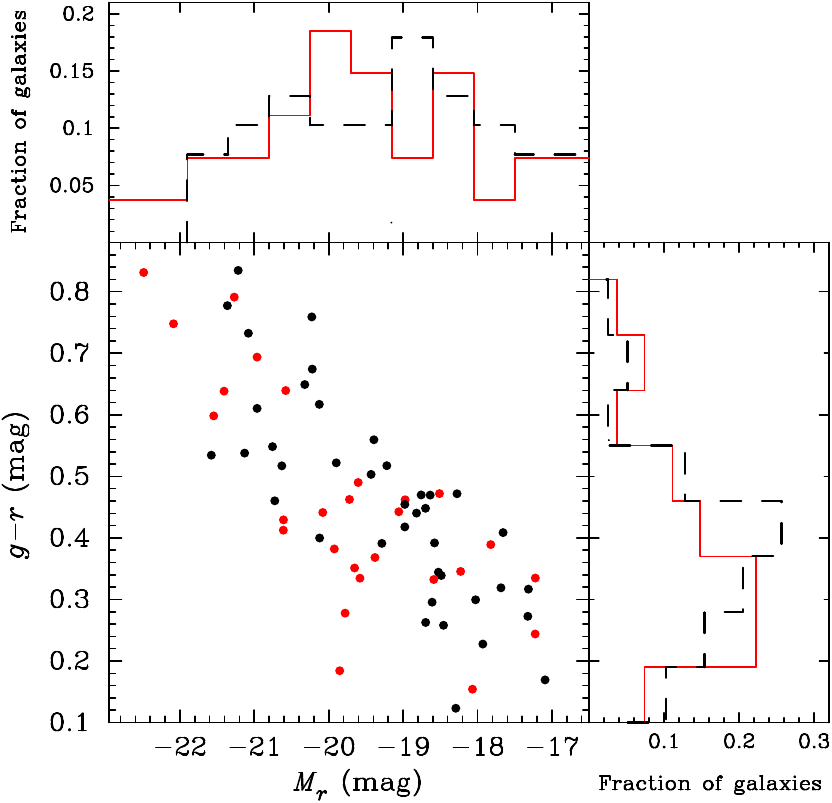}{0.46\textwidth}{(a)}
          \fig{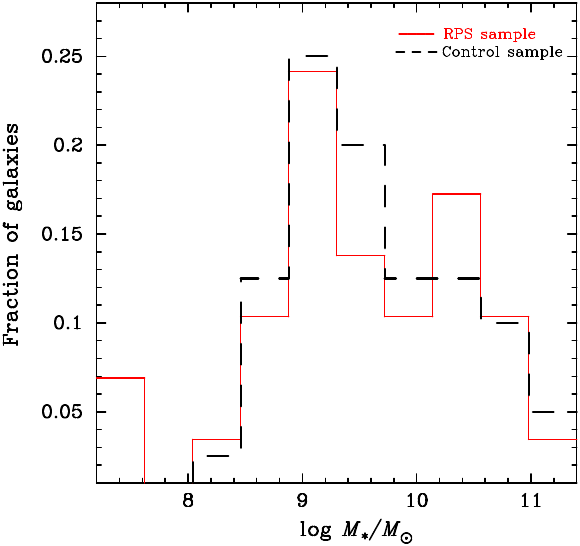}{0.46\textwidth}{(b)}}
  \caption{\small (a) This figure illustrates the distribution of galaxies in the RPS sample (\textit{red solid circles}) and the control sample (\textit{black solid circles}) on the color-magnitude plane. Galaxies in the control sample are selected to mirror the color-magnitude distribution of the RPS galaxies as closely as possible, as evidenced by the color and magnitude histograms (RPS: \textit{red solid}; Control: \textit{black dashed}) shown alongside. (b) This figure shows the stellar mass distributions of the galaxies in the two samples. 50$\%$ of the galaxies in both the samples have a stellar mass of $\ge 10^{9.54} M_{\odot}.$}\label{fig:cmd} 
\end{figure}

\subsection{RPS Sample}
Our RPS sample consists of 29 galaxies belonging to four nearby rich galaxy clusters at $z = 0.015 - 0.024$: Abell~1656 (Coma), Abell~1367 (Leo), Abell~426 (Perseus) and Abell~3627 (Norma).
We adopted the following cluster redshifts (also to be consistent with our previous works): 0.0231 for Coma, 0.0220 for Abell~1367, 0.01753 for Perseus and 0.01605 for Abell~3627. Thus, the scale (kpc per arcsec) and luminosity distance (Mpc) adopted are: 0.466 and 100.7 for Coma, 0.445 and 95.8 for Abell~1367, 0.356 and 76.1 for Perseus, 0.327 and 69.6 for Abell~3627.
These galaxies are typically late-type and undergoing RPS, as evidenced by existing literature on H$\alpha$, UV, X-ray, and radio RPS studies \citep{sun2007,sun2010_A3627,yagi2010_rps, smith2010_rps, fossati2012_rps, yagi2017_rps,gavazzi2018_rps,roberts-parker2020_rps,roberts-parker2021_rps,roberts-parker2022_rps}. Importantly, they all reside within the \textit{Chandra} X-ray footprint\footnote{Three additional galaxies (GMP 3896, GMP 4060, and GMP 2599) undergoing RPS within the galaxy clusters covered in this work were found to have \textit{Chandra} coverage. All of these were non-detections and located far from the \textit{Chandra} observation aimpoint, hence, resulting in high uncertainty in the estimates of flux upper limits. These were therefore excluded from the RPS sample.}, therefore allowing our investigation of their X-ray emission.
We utilized data from the SDSS data release 18 to obtain the photometric properties of the galaxies. The details of galaxies in the RPS sample, along with their properties, are presented in Table \ref{tab:RPS sample basic info.}. In Figure \ref{fig:cmd}(a), we show the distribution of galaxies in the RPS sample on the color-magnitude plane.\footnote{The galaxies ESO 137-001 and ESO 137-002 are excluded from this figure as they fall outside the coverage area of SDSS. \\Their exclusion does not affect our sample selection and results.} As anticipated, a majority of these galaxies exhibit a blue color, signifying their gas-rich nature.

\centerwidetable
\begin{deluxetable}{llllllll}
\small
\setlength{\tabcolsep}{1.5mm}
\tablecaption{Galaxies in the RPS sample and their photometric properties from the SDSS\label{tab:RPS sample basic info.}}
\startdata\\
Galaxy (alias) & RA & Dec & ${v_\text{cluster-centric}}^\text{(f)}$& $r^\text{(a)}$ & $g$-$r^\text{(b)}$ & $M_r^\text{(c)}$&$\text{log } {M_*}^\text{(e)}$\\
&(J2000)&(J2000)&&&&&\\
&h:min:s&$^{\circ}$: $'$: $''$&(km s$^{-1})$&(mag)&(mag)&(mag)&$(M_{\odot})$\\
\hline
                        Abell 1656 (Coma)&&&&&&&\\\hline 
			NGC 4848&12:58:05.60&+28:14:33.53&+263&13.61&0.64&-21.4&10.33\\
			GMP 2910 (Mrk 60, D100)&13:00:09.14&+27:51:59.46 &-1614&15.41&0.49&-19.6&9.27\\
			GMP 2374 (NGC 4911)&13:00:56.08&+27:47:26.98&+1065&12.93&0.75&-22.1&10.90\\
			GMP 2923 (LEDA 126767)&13:00:08.05&+27:46:24.01&+1796&16.79&0.34&-18.2&8.76\\
			GMP 3016&13:00:01.03&+28:04:54.78&+840&17.79&0.34&-17.2&7.50\\
			GMP 3071&12:59:56.12&+27:44:46.72&+2045&16.43&0.33&-18.6&8.91\\
			GMP 3271 (KUG 1257+278)&12:59:39.82&+27:34:35.19&-1923&15.64&0.36&-19.4&9.21\\
			GMP 3779 (Mrk 58)&12:59:05.29&+27:38:40.10&-1496&14.41&0.41&-20.6&9.75\\
			GMP 3816 (NGC 4858)&12:59:02.06&+28:06:56.33&+2500&15.09&0.38&-19.9&9.41\\
			GMP 4017 (NGC 4854)&12:58:47.41&+27:40:28.98&+1452&13.75&0.79&-21.3&10.66\\
			GMP 4156 (NGC 4853)&12:58:35.19&+27:35:47.05&+746&13.47&0.60&-21.5&10.38\\
			GMP 4232&12:58:30.62&+27:33:51.50&+355&17.79&0.24&-17.2&7.40\\
			GMP 2559 (IC 4040)&13:00:37.97&+28:03:26.83&+707&14.94&0.44&-20.1&9.66\\
			GMP 4570&12:57:56.76&+27:59:30.13&-2351&16.95&0.15&-18.1&8.54\\
			GMP 4555 (KUG 1255+283)&12:57:57.83&+28:03:38.53&+1212&15.29&0.47&-19.7&9.76\\
			GMP 4159 (Mrk 56)&12:58:35.33&+27:15:52.92&+426&15.24&0.28&-19.8&9.17\\
			GMP 4281 (LEDA 1827127)&12:58:25.54&+28:07:44.04&+1273&16.04&0.46&-19.0&9.24\\
			GMP 4333 (SDSS J125819.72+280541.4)&12:58:19.73&+28:05:41.28&+242&17.19&0.39&-17.8&8.34\\
			GMP 2059 (NGC 4921)&13:01:26.12&+27:53:09.63&-1449&12.53&0.83&-22.5&11.35\\
			GMP 4351 (LEDA 44437)&12:58:18.62&+27:18:38.88&+517&15.96&0.44&-19.1&9.19\\\hline
                        Abell 1367 (Leo)&&&&&&&\\\hline
			UGC 6697 (CGCG 097-087)&11:43:49.12&+19:58:07.30&+127&14.30&0.43&-20.6&10.16\\
			UGC 6697N (CGCG 097-087N)&11:43:49.71&+19:58:34.07&+942&16.40&0.47&-18.5&8.81\\
			NGC 3860B (MCG+03-30-087)&11:44:47.83&+19:46:24.30&+1868&15.33&0.33&-19.6&9.25\\\hline
			Abell 426 (Perseus)&&&&&&&\\\hline
			LEDA 2191078&3:16:33.92&+42:01:32.59&-1181&14.76&0.35&-19.7&9.57\\
			MCG+07-07-070&3:20:22.02&+41:38:26.59&-1354&14.55&0.19&-19.9&9.54\\
			UGC 2654&3:18:43.10&+42:18:02.09&+534&13.44&0.70&-20.9&10.97\\
			UGC 2665&3:19:27.37&+41:38:07.12&+2495&13.83&0.64&-20.6&10.19\\\hline
			Abell 3627 (Norma) &&&&F475W-F814W$^\text{d}$&&&\\\hline
			ESO 137-001&16:13:27.24&-60:45:50.76&-168&&$\sim$0.5&&9.81\\
			ESO 137-002&16:13:35.73&-60:51:54.71&+876&&$\sim$1.2&&10.54\\
\enddata
\vspace{0.5cm}
\begin{tablenotes}
\small{\item (a): SDSS dereddened $r$-band Petrosian apparent magnitude of galaxies. (b): $g$-$r$ color. (c): Petrosian absolute magnitude in the SDSS $r$-band.(d): ESO 137-001 and ESO 137-002 lie outside the coverage area of SDSS. Therefore, galaxy colors derived from the F475W and F814W filters of the Hubble Space Telescope ACS/WFC (Wide Field Channel of the Advanced Camera for Surveys) instrument are reported. The values are taken from \citet{waldron2023} and \citet{laudari2022} for ESO 137-001 and ESO 137-002, respectively. These closely align with the $g$-$i$ color measurements of other galaxies in the RPS sample acquired from SDSS. Note that the F475W-F814W color of ESO 137-002 is not corrected for intrinsic extinction. (e): Galaxy stellar mass values are sourced from \citet{boselli2022} and have been rescaled to match the cluster distances used in this work. For galaxies in Perseus cluster, the stellar mass values are sourced from \citet{roberts-parker2022_rps}. (f): In calculating the cluster-centric velocities, the heliocentric velocities of galaxies as reported in \citet{boselli2022} or the \citet{ned}, are used; the velocity of ESO 137-001 is taken from \citet{luo2023}. The radial velocity dispersion (km s$^{-1}$) measured for galaxies within $R_{200}$ of Coma and Abell~1367 is 873 and 726, respectively \citep{sohn2020}, 925 for Abell~3627 \citep{woudt2008}, and 1040 for Perseus \citep{aguerri2020}.\\}
\end{tablenotes}
\end{deluxetable}

\subsubsection{\textit{Chandra} X-ray observations and data processing} 
\begin{figure}
\gridline{\fig{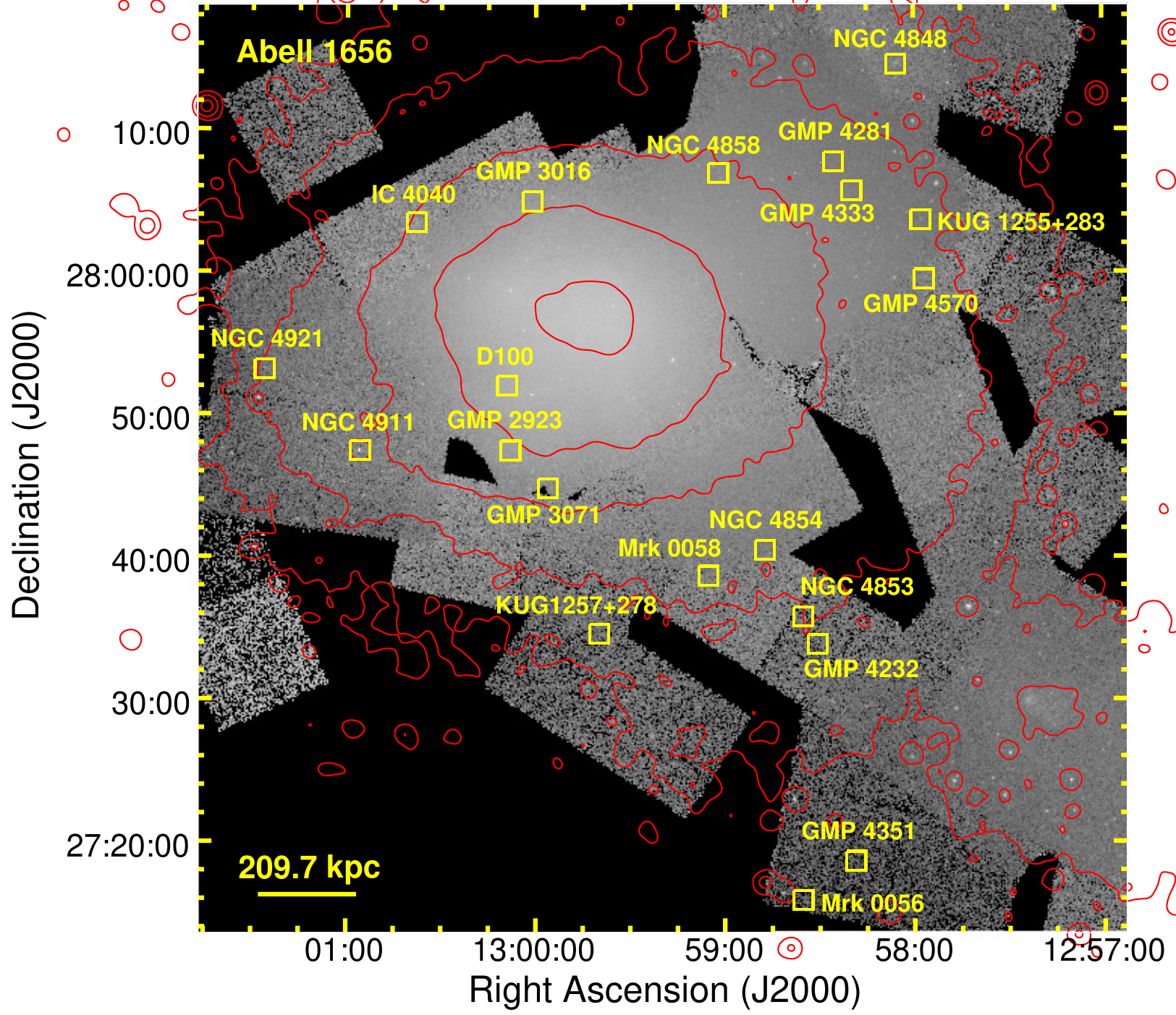}{0.52\textwidth}{(a) Abell 1656 (Coma)}
          \fig{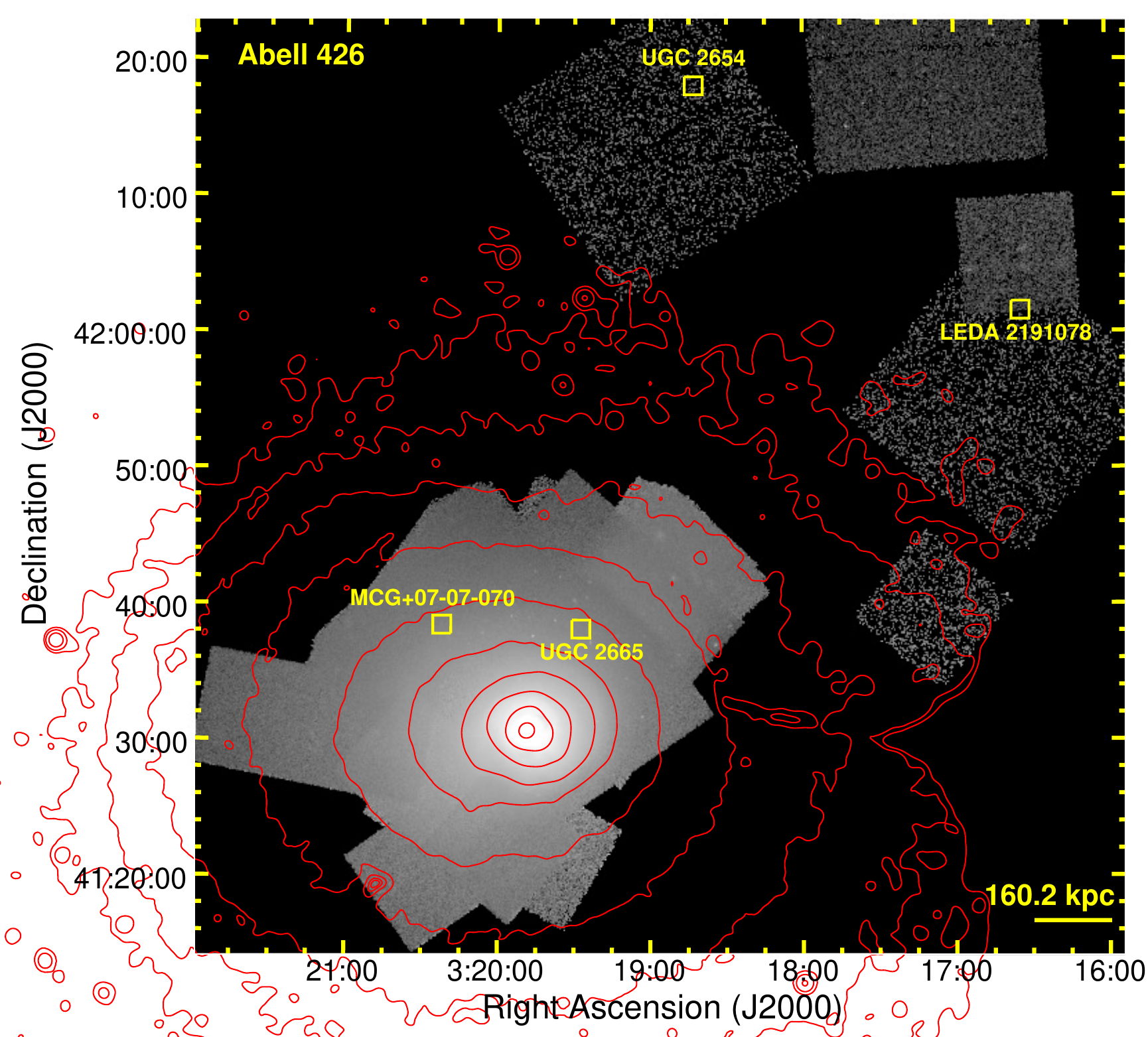}{0.52\textwidth}{(b) Abell 426 (Perseus)}}
\gridline{\fig{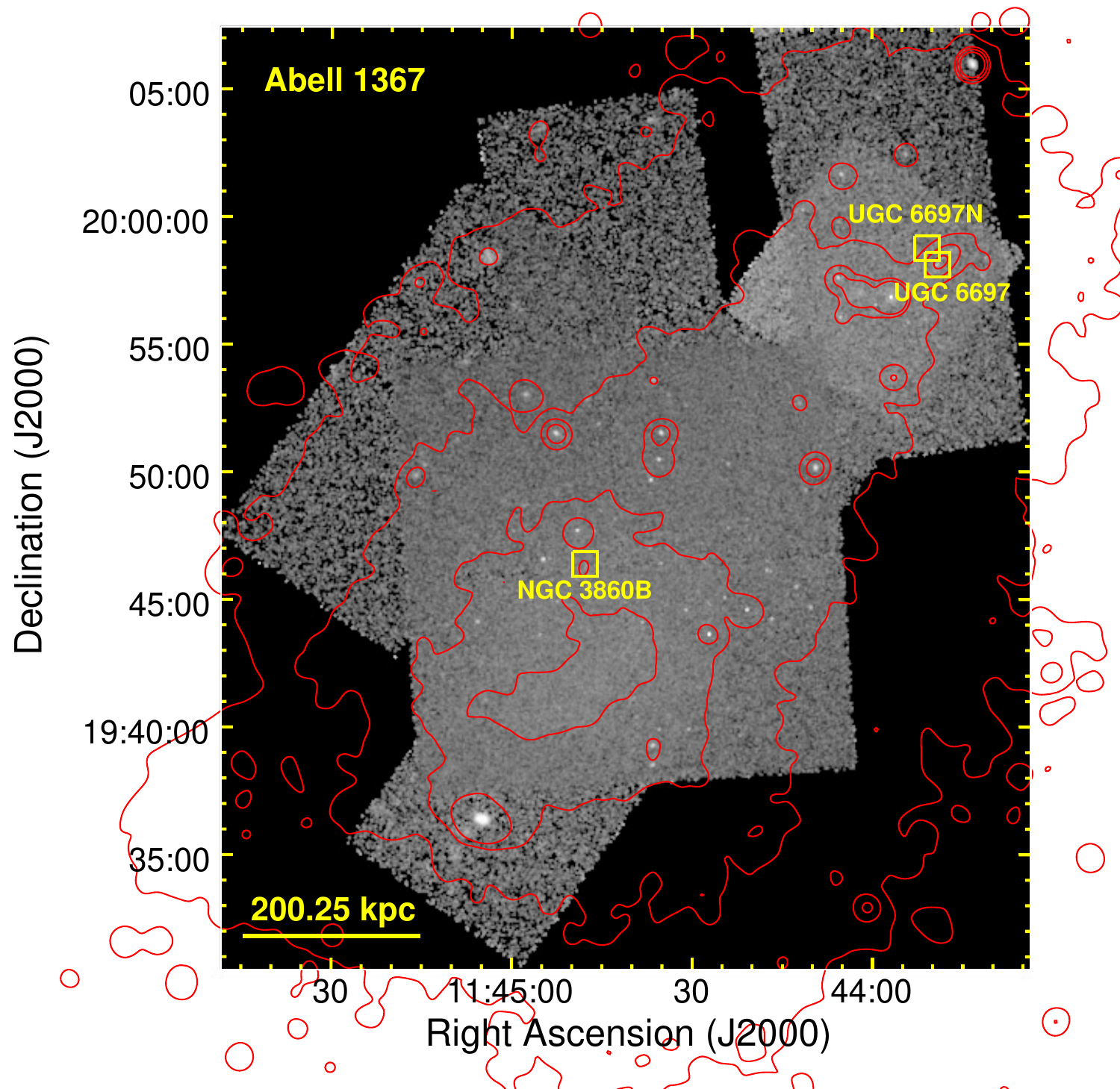}{0.52\textwidth}{(c) Abell 1367 (Leo)}
          \fig{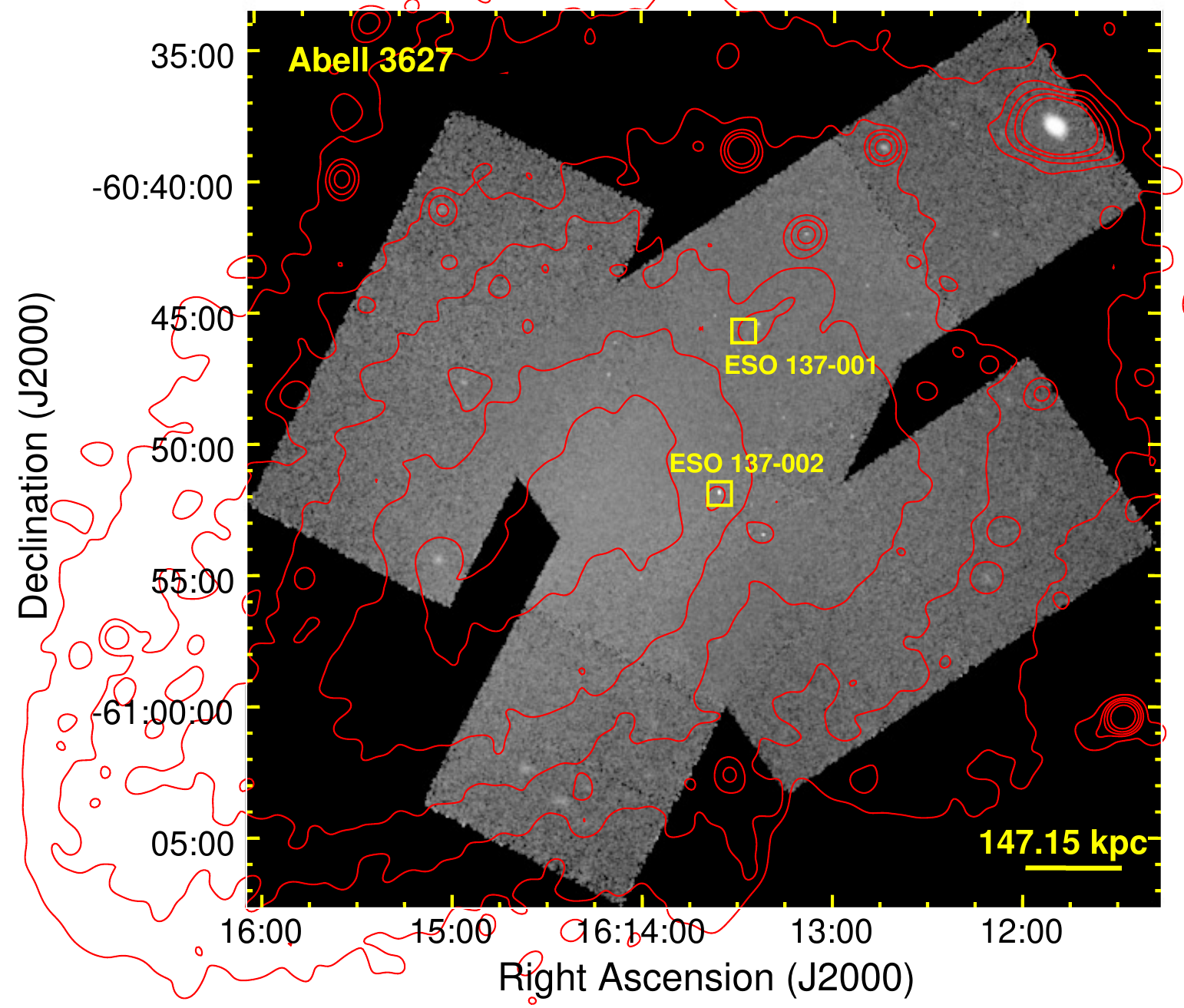}{0.52\textwidth}{(d) Abell 3627 (Norma)}}
  \caption{\small The \textit{Chandra} X-ray flux images of galaxy clusters A1656, A426, A1367, and A3627 are presented in the 2.0--8.0 keV energy band. Each panel features highlighted boxes indicating the positions of cluster member galaxies undergoing RPS. Notably, several massive galaxies ($M_* \ge 10^{10.3} M_{\odot}$), such as NGC 4911, NGC 4921, NGC 4848, and NGC 4853 in panel (a), and ESO 137-002 in panel (d), exhibit distinct signs of hard X-ray point source-like emission. In each panel, the X-ray surface brightness contours in the 0.5--2.0 keV range from \textit{XMM-Newton} are overlaid in \textit{red} to show the extent of the cluster emission. 
  \label{fig:xraycl_rps}}  
\end{figure}

The \textit{Chandra} ACIS X-ray observations for all galaxies in the RPS sample are detailed in Table \ref{tab:X-ray observations of RPS sample}. Data from these observations were retrieved from the Chandra Data Archive\footnote{\url{https://cda.harvard.edu/chaser/}} and analysed using the \textit{Chandra} Interactive Analysis of Observations (CIAO) software version 4.14 and CALDB version 4.10.4. The light curves of the resulting \textit{level=2} event files were filtered for soft proton flares. Subsequently, we employed the CIAO script \textit{merge$\_$obs} to combine all \textit{Chandra} observations, generating X-ray mosaics of the four galaxy clusters used in the RPS sample selection. We then ran the CIAO task \textit{wavdetect} with wavelet \textit{scales=2,4} and \textit{ellsigma=5} on each of the mosaics to detect X-ray point sources associated with the galaxies in the RPS sample in two energy bands: 0.5--2.0 keV (soft) and 2.0--8.0 keV (hard). To provide visual context, Figure \ref{fig:xraycl_rps} displays the X-ray mosaics of the four galaxy clusters studied in this work, highlighting the positions of the member RPS galaxies in the 2.0–8.0 keV energy band.\vspace{0.2cm}\\

\subsubsection{X-ray Fluxes and Luminosities}
To estimate the fluxes of the detected point sources, we employed spectral fitting whenever possible based on the available number of counts. In many instances, there are multiple X-ray observations for a given source (see Table \ref{tab:X-ray observations of RPS sample}). We determined the radius of the circle enclosing $90\%$ of the point spread function (PSF) at an effective energy of 2 keV at the location\footnote{The centre of the source was confirmed by comparing it with the SDSS \textit{i}-band image in the case of detections.} of each source for each observation. Since this radius varies from observation to observation due to different aim points, even for a fixed sky position and effective energy, we adopted the maximum radius from among the different observations. This maximum radius was employed to define the circular region for spectral extraction for each source. Spectra and responses for each source were extracted using the task \textit{specextract}. In case of multiple observations for a given source, spectra and responses were combined using \textit{combine$\_$spectra}. A background spectrum was extracted from within an annulus around each source having a size five times the radius of the source. The spectra were appropriately grouped using the FTOOL \textit{grppha} and analysed using XSPEC version 12.13.1. Each source spectrum was fitted in the energy range 0.5--7.0 keV using an absorbed powerlaw model \textit{tbabs*powerlaw}. Here, \textit{tbabs} is the Tuebingen-Boulder interstellar medium absorption model and uses the ISM elemental abundances as given by \citet{wilms2000}; The total hydrogen column density along the line of sight to the source ($\textit{N}_H$) was chosen based on the method of \citet{willingale2013}\footnote{The tool \url{https://www.swift.ac.uk/analysis/nhtot/} returns the neutral, molecular, and total galactic column density of hydrogen along a given line of sight.}. The component \textit{powerlaw}\footnote{The \textit{powerlaw} model has the form $A(E)=KE^{-\alpha}$, where $\alpha$ is the dimensionless photon index and $K$ is the model normalization in units of photons$/$keV$/$cm$^2/$s at 1 keV.} describes the photon spectrum of the point source under consideration; the photon index ($\alpha$) of the \textit{powerlaw} was fixed at 1.7. In cases where there was some contamination from the surrounding intracluster gas, we added an \textit{apec} component \citep{smith2001} to model this contribution. We also tested for the possibility of the X-ray emission from the point source being affected due to intrinsic absorption by adding another absorption model component. Only ESO 137-002 was found to exhibit significant intrinsic absorption (see Table \ref{tab:X-ray observations of RPS sample} for details; and see \citealt{sun2010_A3627} and \citealt{zhang2013}). The unabsorbed X-ray fluxes and luminosities were estimated by convolving the best-fitting \textit{powerlaw} model component with the XSPEC models \textit{cflux} and \textit{clumin} respectively. These are listed in Table \ref{tab:X-ray observations of RPS sample}.

If a spectral analysis was not feasible due to a low number of counts, we used the CIAO script \textit{srcflux} to estimate the unabsorbed model fluxes of the sources. The model employed for this estimation is the same as that used in the spectral analysis. For calculating the source count rate, the script utilized a circle centered on each specified source location\footnote{For non-detections, the centre from the SDSS \textit{i}-band image was selected as the source centre for estimating flux upper limits.
} (given in Table \ref{tab:RPS sample basic info.}) with radius equivalent to that required to enclose 90$\%$ of the PSF at 2.0 keV. The determination of the background rate involved the script employing an annular background region centered on the source location, with an inner radius matching the size of the source radius and an outer radius set at five times the inner radius. The resulting flux and luminosity values are presented in Table \ref{tab:X-ray observations of RPS sample}.

Several galaxies in the RPS sample are covered by \textit{Chandra}, yet no point source was detected for them (via \textit{wavdetect} or visually). In this instance, we calculated the 5-$\sigma$ upper limits for their X-ray fluxes and luminosities. To do this, we estimated the number of counts in a background region chosen around the undetected source and scaled this value to its assumed size -- a circular region enclosing 90$\%$ of the PSF at 2 keV. The count rate at the 5-$\sigma$ level above the background within this putative point source was then converted into flux in the required energy range using the CIAO script \textit{modelflux}, employing the same absorbed powerlaw model as used above.  The upper limits for X-ray fluxes and luminosities for non-detections (12 out of 29 galaxies in the RPS sample) are included in Table \ref{tab:X-ray observations of RPS sample}.

\startlongtable
\centerwidetable
\begin{deluxetable}{llllllll}
\small
\setlength{\tabcolsep}{0.5mm}
\tablecaption{\textit{Chandra} X-ray observations, flux, and luminosity of the central point source in galaxies in the RPS sample \label{tab:X-ray observations of RPS sample}}
\startdata
Galaxy & ObsID (Clean time) & $F_X$ (2--10 keV) &$L_X$ (2--10 keV)&$F_X$ (0.5--10 keV)&$L_X$ (0.5--10 keV)&Method$^{\text{(a)}}$\\
& (ks)  & (erg s$^{-1}$ cm$^{-2}$) &(erg s$^{-1}$) &(erg s$^{-1}$ cm$^{-2}$)&(erg s$^{-1}$)& \\ \hline
Abell 1656 (Coma)&&&&&\\\hline
NGC 4848&\enspace 8188 (28.66), 18234 (19.79),&$4.3_{-1.6}^{+1.9}\times 10^{-15}$&$5.2_{-1.9}^{+2.3}\times10^{39}$&$9.5\pm{0.2}\times10^{-15}$&$1.2^{+0.02}_{-0.07}\times 10^{40}$&\textit{powerlaw} assumed\\
&20049 (19.79), 20050 (13.87),&$4.2\pm{0.2}\times 10^{-15}$&$5.1\pm{2.2}\times10^{39}$&$6.5\pm{0.3}\times10^{-15}$&$7.9\pm{3.4}\times 10^{39}$&spectral analysis\\
&20051 (14.85), 20052 (23.73) &&&&\\
GMP 2910$^*$ &\enspace 9714 (29.65), 13993 (39.56),&$0.5_{-0.5}^{+2.5}\times10^{-15}$&$0.6_{-0.6}^{+3.0}\times10^{39}$&$5.6\pm{2.3}\times10^{-15}$&$6.8\pm{2.8}\times10^{39}$&\textit{powerlaw} assumed\\
&13994 (81.99), 13995 (62.98),&$8.0_{-7.9}^{+7.6}\times10^{-15}$&$9.7^{+9.2}_{-9.6}\times10^{39}$&$3.0_{-0.7}^{+1.6}\times10^{-15}$&$3.5_{-0.7}^{+2.0}\times10^{39}$&spectral analysis\\
			&13996 (123.06),14406 (24.76),&&&&\\
			&14410 (78.53), 14411 (33.64),&&&&\\
			&14415 (34.53)&&&&\\
			GMP 2374 &\enspace 2941 (62.91)&$1.6\pm{0.1}\times 10^{-13}$&$1.9\pm{0.1}\times 10^{41}$&$2.5\pm{0.1}\times 10^{-13}$&$3.0\pm{0.1}\times 10^{41}$&\textit{powerlaw} assumed\\
            &&$1.4\pm{0.1}\times 10^{-13}$&$1.7\pm{0.1}\times 10^{41}$&$2.2\pm{0.1}\times 10^{-13}$&$2.6\pm{0.1}\times 10^{41}$&spectral analysis\\
			GMP 2923&13993 (39.56), 13994 (81.99),&$<3.9\times 10^{-14}$&$<4.8\times10^{40}$&$<6.1\times10^{-14}$&$<7.4\times10^{40}$&upper limit\\
			&14406 (24.76), 14410 (78.53),&&&&\\
			&14415 (34.53)&&&&\\
			GMP 3016&13995 (62.98), 13996 (123.06),&$<2.9\times10^{-14}$&$<3.5\times10^{40}$&$<4.5\times10^{-14}$&$<5.5\times10^{40}$&upper limit\\
			&14406 (24.76), 14415 (34.53),&&&&\\
			&18235 (29.67)&&&&\\
			GMP 3071&13994 (81.99)&$<9.1\times10^{-14}$&$<1.1\times10^{41}$&$<1.4\times10^{-13}$&$<1.7\times10^{41}$&upper limit\\
			GMP 3271&18237 (9.93), 19911 (9.91),&$<2.2\times10^{-14}$&$<2.7\times10^{40}$&$<3.4\times10^{-14}$&$<4.1\times10^{40}$&upper limit\\
			&19912 (9.92)&&&&\\
			GMP 3779&same as GMP 3271&$<1.8\times10^{-15}$&$<2.2\times10^{39}$&$<2.9\times10^{-15}$&$<3.5\times10^{39}$&upper limit\\
			GMP 3816 &18271 (54.35), 18272 (19.82),&$2.3\pm{1.0}\times 10^{-15}$&$2.8\pm{1.2}\times 10^{40}$&$1.1\pm{0.1}\times 10^{-14}$&$1.3\pm{0.1}\times 10^{40}$&\textit{powerlaw} assumed\\
            &18273 (28.32), 18274 (46.46),&$3.5\pm{1.2} \times 10^{-15}$&$4.2\pm{1.4}\times 10^{39}$&$5.4\pm{1.8}\times 10^{-15}$&$6.5\pm{2.2}\times 10^{39}$&spectral analysis\\
			&18275 (49.43), 18276 (84.18),&&&&\\
			&18761 (47.45), 18791 (34.62),&&&&\\
			&18792 (21.25), 18793 (76.94),&&&&\\
			&18794 (29.30), 18795 (27.78),&&&&\\
			&18796 (39.56), 18797 (54.35),&&&&\\
			&18798 (12.91), 19998 (31.65),&&&&\\
			&20010 (58.30), 20011 (44.49),&&&&\\
			&20027 (20.81), 20028 (43.51),&&&&\\
			&20029 (21.79), 20030 (33.15),&&&&\\
			&20031 (10.93), 20037 (16.86),&&&&\\
			&20038 (49.92), 20039 (21.89)&&&&\\
			GMP 4017&18237 (9.93), 19911 (9.91),&$2.9^{+4.5}_{-2.8}\times10^{-15}$&$3.5_{-3.4}^{+5.4}\times10^{39}$&$5.7_{-3.2}^{+4.3}\times10^{-15}$&$6.9_{-3.9}^{+5.2}\times10^{39}$&\textit{powerlaw} assumed\\
			&19912 (9.92)&&&&\\
			GMP 4156 &same as GMP 4017&$1.6_{-0.8}^{+1.0}\times 10^{-14}$&$1.9_{-0.9}^{+1.2}\times 10^{40}$&$2.8_{-0.8}^{+1.0}\times 10^{-14}$&$3.4_{-1.0}^{+1.2}\times 10^{40}$&\textit{powerlaw} assumed\\
            &&$1.5\pm{0.6}\times 10^{-14}$&$1.8\pm{0.9}\times 10^{40}$&$2.4\pm{1.2}\times 10^{-14}$&$2.8\pm{1.5}\times 10^{40}$&spectral analysis\\
			GMP 4232&same as GMP 4017&$<2.6\times10^{-14}$&$<3.2\times10^{40}$&$<4.0\times10^{-14}$&$<4.9\times10^{40}$&upper limit\\
			GMP 2559&18235  (29.67)&$<2.1\times10^{-15}$&$<2.5\times10^{39}$&$<3.3\times10^{-15}$&$<4.0\times10^{39}$&upper limit\\
			GMP 4570&18236 (9.94), 19909 (9.92),&$<8.1\times10^{-15}$&$<9.8\times10^{39}$&$<1.3\times10^{-14}$&$<1.6\times10^{40}$&upper limit\\
			&19910 (9.91)&&&&\\
			GMP 4555&18236 (9.94), 18271 (54.35),&$9.9\pm{1.7}\times 10^{-15}$&$1.2\pm{0.2}\times 10^{40}$&$2.0\pm{0.2}\times 10^{-14}$&$2.4\pm{0.3}\times 10^{40}$&\textit{powerlaw} assumed\\
            &18272 (19.82), 18274 (46.4),&$9.9\pm{2.1}\times 10^{-15}$&$1.2\pm{0.3}\times 10^{40}$&$1.5\pm{0.3}\times 10^{-14}$&$1.8\pm{0.4}\times 10^{40}$&spectral analysis\\
			&18275 (49.43),	18276 (84.18),&&&&\\
			&18761 (47.45), 18791 (34.62),&&&&\\
			&18792 (21.25), 18793 (76.94),&&&&\\
			&18794 (29.30), 18795 (27.78),&&&&\\
			&18796 (39.56), 18797 (54.35),&&&&\\
			&18798 (12.91), 19909 (9.92),&&&&\\
			&19910 (9.91), 19998 (31.65),&&&&\\
			&20027 (20.81), 20028 (43.51),&&&&\\
			&20029 (21.79), 20030 (33.15),&&&&\\
			&20031 (10.93), 20037 (16.86),&&&&\\
			&20038 (49.92), 20039 (21.89)&&&&\\
			GMP 4159 &\enspace 4724 (59.68)&$8.0^{+8.5}_{-6.7}\times 10^{-15}$&$9.7^{+10.3}_{-8.2}\times 10^{39}$&$13.4_{-4.5}^{+6.5}\times 10^{-15}$&$16.3^{+7.8}_{-5.5}\times 10^{39}$&\textit{powerlaw} assumed\\
            &&$1.8^{+5.1}_{-1.8}\times 10^{-15}$&$2.3^{+6.0}_{-2.3}\times 10^{39}$&$2.8_{-2.8}^{+7.8}\times 10^{-15}$&$3.3^{+9.5}_{-3.3}\times 10^{39}$&spectral analysis\\
			GMP 4281 &18236 (9.94), 18271 (54.35),&$<2.1\times10^{-15}$&$<2.5\times10^{39}$&$<3.2\times10^{-15}$&$<3.9\times10^{39}$&upper limit\\
			&18273 (28.32), 18274 (46.46),&&&&\\
			&18275 (49.43), 18276 (84.18),&&&&\\
			&18791 (34.62), 18792 (21.25),&&&&\\
			&18793 (76.94), 18794 (29.30),&&&&\\
			&18795 (27.78), 18796 (39.56),&&&&\\
			&18797 (54.35), 18798 (12.91),&&&&\\
			&19909 (9.92), 19910 (9.91),&&&&\\
			&19998 (31.65), 20010 (58.30),&&&&\\
			&20011 (44.49), 20028 (43.51),&&&&\\
			&20030 (33.15), 20037 (16.86),&&&&\\
			&20038 (49.92), 20039 (21.89)&&&&\\
			GMP 4333 &same as GMP 4281&$<2.2\times10^{-15}$&$<2.7\times10^{39}$&$<3.4\times10^{-15}$&$<4.1\times10^{39}$&upper limit\\
			GMP 2059 &\enspace 2941 (62.91)&$1.5\pm{0.5}\times 10^{-14}$&$1.8\pm{0.6}\times 10^{40}$&$1.9\pm{0.4}\times 10^{-14}$&$2.3\pm{0.5}\times 10^{40}$&\textit{powerlaw} assumed\\
            &&$1.1\pm{0.3}\times 10^{-14}$&$1.3\pm{0.3}\times 10^{40}$&$1.7\pm{0.4}\times 10^{-14}$&$2.0\pm{0.5}\times 10^{40}$&spectral analysis\\
			GMP 4351 &\enspace 4724 (59.68)&$3.9_{-3.8}^{+4.9}\times10^{-15}$&$4.7^{+5.9}_{-4.6}\times10^{39}$&$7.3_{-3.0}^{+3.8}\times10^{-15}$&$8.9^{+4.6}_{-3.7}\times10^{39}$&\textit{powerlaw} assumed\\\\\hline
			Abell 1367 (Leo)&&&&&\\\hline
			UGC 6697 &\enspace 4189 (39.58)&$1.5_{-0.4}^{+0.6}\times 10^{-14}$&$1.7\pm{0.5}\times 10^{40}$&$2.1\pm{0.3}\times 10^{-14}$&$2.3\pm{0.4}\times 10^{40}$&\textit{powerlaw} assumed\\
            &&$1.2\pm{0.2}\times 10^{-14}$&$1.3\pm{0.3}\times 10^{40}$&$1.9\pm{0.4}\times 10^{-14}$&$2.1\pm{0.4} \times 10^{40}$&spectral analysis\\
			UGC 6697N &same as UGC 6697&$2.3_{-1.6}^{+2.8}\times10^{-15}$&$2.5^{+3.1}_{-1.8}\times10^{39}$&$1.4_{-0.9}^{+1.3}\times10^{-15}$&$1.5^{+1.5}_{-1.0}\times10^{39}$&\textit{powerlaw} assumed\\
			NGC 3860B &\enspace \enspace 514 (36.98), 17199 (37.93),&$9.4_{-9.4}^{+13.0}\times 10^{-16}$&$1.0^{+1.5}_{-1.0}\times10^{39}$&$3.1_{-1.3}^{+1.4}\times10^{-15}$&$3.4\pm{1.5}\times10^{39}$&\textit{powerlaw} assumed\\
			&17200 (39.56), 17201 (61.27),&&&&\\
			&17589 (24.76), 17590 (36.59),&&&&\\
			&17591 (34.92), 17592 (25.73),&&&&\\
			&18704 (27.03), 18705 (23.76),&&&&\\
			&18755 (48.44)&&&&\\\\\hline
			Abell 426 (Perseus)&&&&&\\\hline
			LEDA 2191078&17279 (4.71)&$<1.2\times10^{-14}$&$<8.3\times10^{39}$&$<1.9\times10^{-14}$&$<1.3\times10^{40}$&upper limit\\
			MCG+07-07-070&\enspace 4946 (23.66), \enspace 4947 (29.79),&$7.9_{-7.9}^{+32.0}\times10^{-16}$&$5.5_{-5.5}^{+22.2}\times10^{38}$&$1.4\pm{0.4}\times10^{-14}$&$9.7\pm{2.8}\times10^{39}$&\textit{powerlaw} assumed\\
			&\enspace 4948 (118.61), 4949 (29.38),&&&&\\
			&\enspace 4950 (96.92), \enspace 4951 (96.12),&&&&\\
			&\enspace 4952 (164.24), 4953 (30.08),&&&&\\
			&11713 (112.24), 11714 (91.99),&&&&\\
			&12036 (47.92)&&&&\\
			UGC 2654&17276 (5.02)&$1.2^{+2.4}_{-1.1}\times10^{-14}$&$8.3^{+16.6}_{-7.6}\times10^{39}$&$8.1^{+17.4}_{-7.9}\times10^{-15}$&$5.6^{+2.1}_{-5.5}\times10^{39}$&\textit{powerlaw} assumed\\
			UGC 2665&\enspace 3209 (95.77), \enspace 4289 (95.41),&$<1.0\times10^{-14}$&$<6.9\times10^{39}$&$<1.6\times10^{-14}$&$<1.1\times10^{40}$&upper limit\\
			&\enspace 4946 (23.66), \enspace 4947 (29.79),&&&&\\
			&\enspace 4948 (118.61), 4949 (29.38),&&&&\\
			&\enspace 4950 (96.92), \enspace 4951 (96.12),&&&&\\
			&\enspace 4952 (164.24), 4953 (30.08),&&&&\\.
			&\enspace 6139 (56.43), \enspace 6145 (85.00),&&&&\\
			&\enspace 6146 (47.13), 11713 (112.24),&&&&\\
			&11714 (91.99), 12025 (17.93),&&&&\\
			&12033 (18.89), 12036 (47.92)&&&&\\\hline
			Abell 3627 (Norma) &&&&&\\\hline
			ESO 137-001&\enspace 9518 (140.04)&$1.4_{-0.9}^{+1.2}\times 10^{-15}$&$8.1_{-5.3}^{+6.6}\times10^{38}$&$6.0\pm{1.5}\times10^{-15}$&$3.5\pm{0.8}\times10^{39}$&\textit{powerlaw} assumed\\
            &&$1.6\pm{0.9}\times 10^{-15}$&$9\pm{5}\times 10^{38}$&$2.4\pm{1.4}\times 10^{-15}$&$1.4\pm{0.8}\times 10^{39}$&spectral analysis\\
			ESO 137-002$^\text{(b)}$&12950 (89.85)&$1.9\pm{0.1}\times10^{-13}$&$1.1\pm{0.05}\times10^{41}$&$1.2\pm{0.1}\times10^{-13}$&$7.0\pm{0.5}\times10^{40}$&\textit{powerlaw} assumed\\&&$1.5\pm{0.2}\times10^{-12}$&$7.8\pm{1.2}\times10^{41}$&$2.2\pm{0.3}\times10^{-12}$&$1.2\pm{0.2}\times10^{42}$&spectral analysis\\          
\enddata
\end{deluxetable}
\begin{tablenotes}
       \small{\item (a): This column reports the method used to estimate the unabsorbed fluxes and luminosities of the presumed AGN in each galaxy in the RPS sample. All errors are quoted at 90 per cent confidence level.  \\For detections, a \textit{powerlaw} model with a photon index value of $1.7$ was assumed (Method: \textit{powerlaw} assumed). \\For 10 galaxies in the sample, the number of available counts made it possible to perform a spectral analysis. We also report the flux and luminosity values of the central AGN component for these galaxies (Method: spectral analysis). $^*$In GMP 2910, however, the spectral analysis of the central region was not consistent with the presence of a non-thermal \textit{powerlaw} component. The flux and luminosity values reported for GMP 2910 are for the thermal gas component; \\For non-detections, we report the 5-$\sigma$ upper limits for the X-ray fluxes and luminosities (Method: upper limit). 
        \item (b): For ESO 137-002, the `\textit{powerlaw} assumed' and `spectral analysis' methods result in dissimilar flux/luminosity values because the `spectral analysis' method takes into account the intrinsic absorption $(\sim4\times10^{23} \text{cm}^{-2})$ in the AGN source whereas the `\textit{powerlaw} assumed' model does not.}
    \end{tablenotes}

\normalsize

\subsection{Control Sample}
We utilized the galaxy environment catalogs by \citet{tempel2014}\footnote{The catalogs can be accessed at \url{http://cdsarc.u-strasbg.fr/viz-bin/qcat?J/A+A/566/A1}} to form our control sample of galaxies. The authors constructed flux-limited and volume-limited galaxy group and cluster catalogs from the spectroscopic sample of the SDSS data release 10 galaxies, employing a modified friends-of-friends method with variable linking length. Since our objective is to explore whether the enhancement of AGN activity is linked to RPS in cluster galaxies, a comparison with X-ray AGN activity in galaxies not subject to RPS (non-RPS) would serve as an optimal control sample.
We note that forming a significantly sized sample of non-RPS late-type spiral galaxies in galaxy clusters poses a significant challenge. Interactions with the dense and hot intracluster medium mean that galaxies are almost never entirely free of RPS. Even in the outskirts of clusters where the pressure or effect of RPS may be lower compared to inner regions, it still exists. Late-type spiral galaxies, with their extended gas disks, are particularly susceptible to this environmental effect. As a result, non-RPS late-type spirals are a rare occurrence within clusters, as they often undergo some degree of gas stripping or morphological transformation due to their environment. We therefore selected all spiral galaxies with SDSS spectroscopic CLASS=GALAXY from the \citet{tempel2014} catalogs without cluster or group membership (hence fairly non-RPS), aligning them with similar redshift ($0.007 \leq z \leq 0.04$), color ($0.12 \leq g-r \leq 0.83$), and absolute magnitude range ($-22.5 \leq M_r \leq -17.1$) as the galaxies in our RPS sample to ensure a fair comparison between the two samples. These selection cuts resulted in an initial sample of 17979 galaxies. To assess the Chandra ACIS coverage of these galaxies, we conducted a query in the Chandra Master Catalog\footnote{\url{https://heasarc.gsfc.nasa.gov/W3Browse/all/chanmaster.html}}, searching for X-ray observations within 8 arcmin of the equatorial coordinates of each of the 17979 galaxies. In constructing our final control sample, we limited our selection to observations with an exposure time exceeding 25 ks, maintaining a maximum search offset of 5 arcmin and 4 arcmin for ACIS-I and ACIS-S, respectively. These specific selection criteria were carefully chosen to ensure the accuracy of deriving flux values for detections and establishing confident upper limits for non-detections, utilizing the same methods applied to galaxies in the RPS sample. The final control sample comprises 40 sources, including 23 detections and 17 non-detections. This selection is designed to closely match the color-magnitude distribution of galaxies in the RPS sample as shown in Figure \ref{fig:cmd}(a). A visual examination of the SDSS images of these galaxies indicated that none of them are undergoing mergers. Additionally, the stellar masses of galaxies in our control sample span a range from $10^{8.4}$ to $10^{11.1}$ M$_{\odot}$\footnote{The values are from the SDSS Wisconsin PCA-based Stellar Masses and Velocity Dispersions Value Added Catalog.}, similar to those observed in the RPS sample (see Table \ref{tab:RPS sample basic info.}), allowing for a robust comparison between the two samples. The stellar mass distributions of the galaxies in both samples are shown in Figure \ref{fig:cmd}(b). We note that differences in stellar mass estimates across studies can arise due to various factors such as variations in the observational data used (e.g., different bands, resolution, sensitivity) and variations in the methods and algorithms employed for analysis (e.g., different calibration relations, underlying assumptions, utilized models). To achieve parity among estimates, spectral energy distribution (SED) fitting can be a powerful technique to provide more comprehensive and robust estimates of stellar masses compared to methods that rely on single-band photometry or simple scaling relations because it takes into account a wide range of observational data. This is, however, beyond the scope of our current work.

 \begin{figure}
 \gridline{\fig{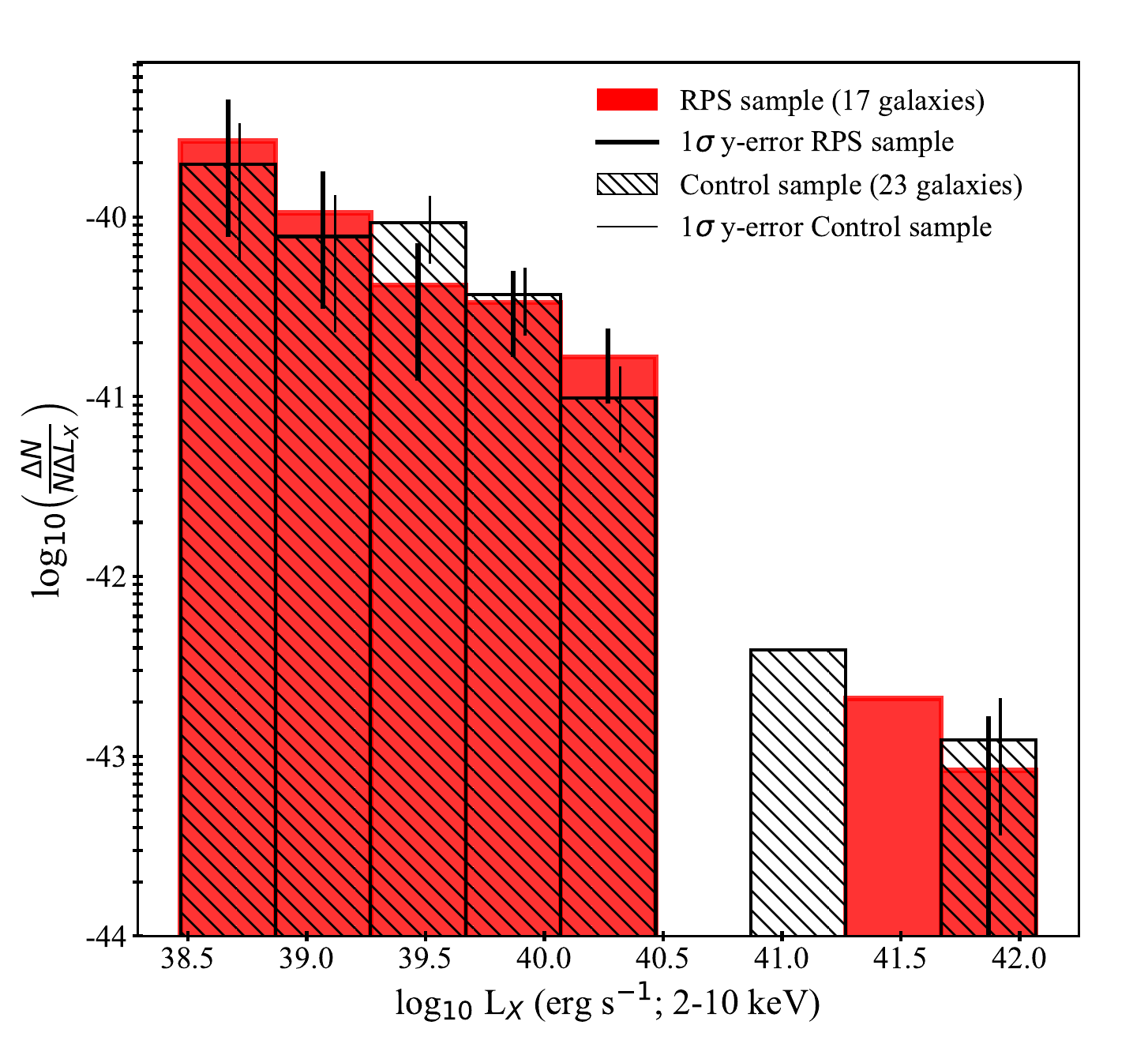}{0.52\textwidth}{(a)}
          \fig{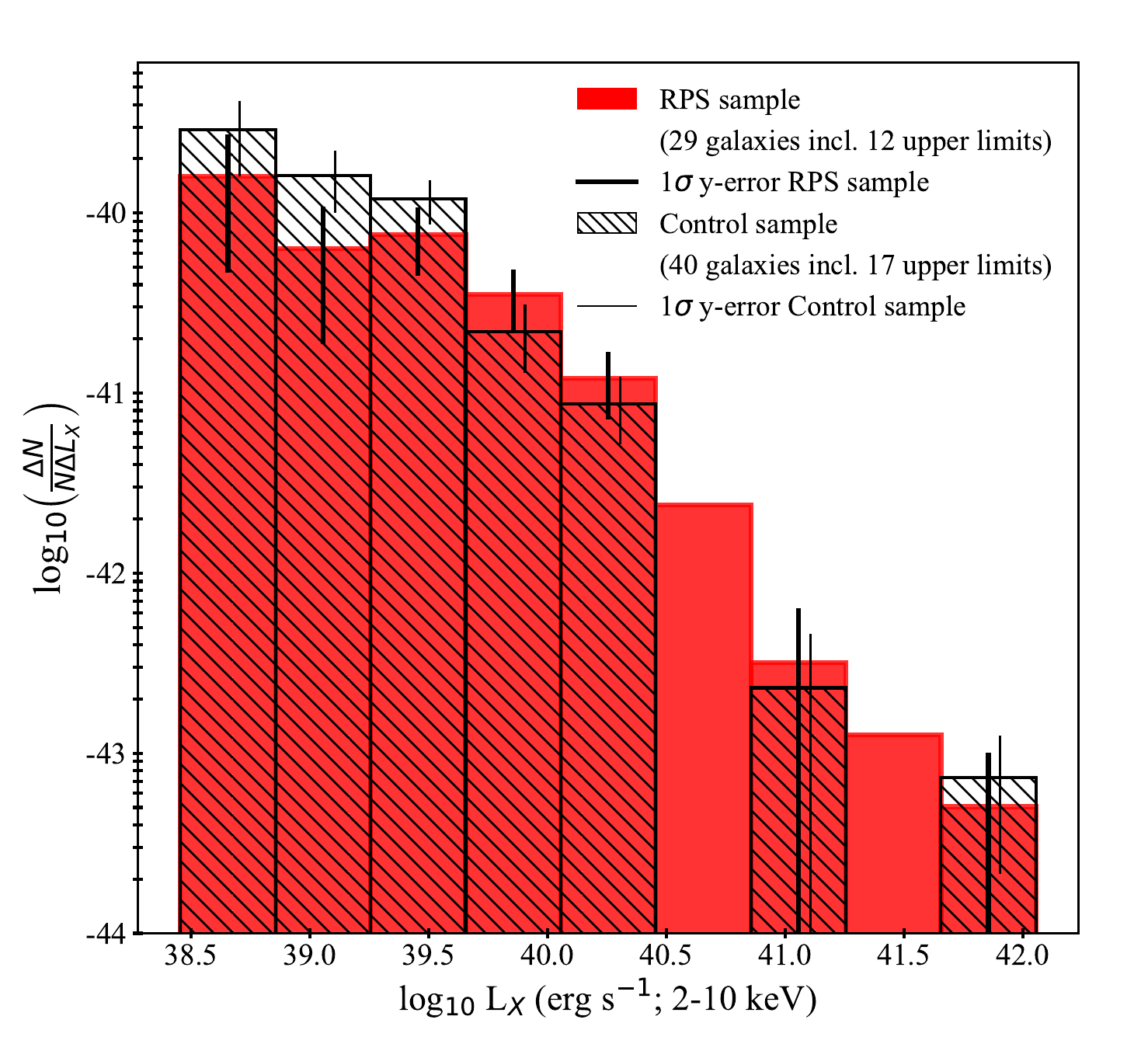}{0.52\textwidth}{(b)}}
    \caption{\small This figure compares the 2--10 keV X-ray luminosity distributions of the central point source in galaxies in the RPS (\textit{solid red}) and Control (\textit{hatched black}) samples. \textit{Thick black} and \textit{thin black} vertical lines represent the Poissonian errors on the histograms of the RPS and Control samples, respectively. The distributions in  panel (b) include the X-ray luminosity upper limits for galaxies with undetected central sources.
    \label{fig:hist_lum}}
    \end{figure}
\section{Results and Discussion}\label{sec:results and discussion}
\subsection{Does RPS enhance X-ray AGN activity?}
In Figure \ref{fig:hist_lum}(a), we present a comparative analysis of the 2--10 keV X-ray luminosity distributions of the detected central sources in galaxies in the RPS and Control samples. Notably, all galaxies (except ESO 137-002) within the RPS sample exhibit central X-ray luminosities below $10^{42}$ erg s$^{-1}$, a commonly used threshold for classifying sources as unambiguous AGN in X-ray studies (e.g., \citealt{bauer2004,alexander2005,xue2011,xue2016,luo2017}). This trend persists even when considering upper limits on luminosity for galaxies with undetected central sources, as illustrated in Figure \ref{fig:hist_lum}(b). These findings suggest a lack of evidence supporting an enhancement in X-ray AGN activity due to RPS phenomena in the observed galaxies.
Additionally, a comparison of the X-ray luminosity distributions of central sources in galaxies between the RPS and Control samples, as depicted in Figure \ref{fig:hist_lum}(a) and Figure \ref{fig:hist_lum}(b), reveals no noticeable difference, further indicating the absence of an increase in X-ray AGN activity from RPS. A two-sample Kolmogorov-Smirnov test run between the nuclear X-ray luminosities in the RPS and Control samples also supports this conclusion, with a p-value of 0.97
with only detections included. At a significance level of $\alpha = 0.05$, this result indicates a statistically non-significant difference between the two distributions.

\subsection{Comparison with previous studies on RPS-AGN connection}
\label{ref:subsec_compare_previous_studies}
In the study by \citet{poggianti2017_nature}, fresh insights into the relationship between AGN activity and RPS in cluster galaxies emerged. The authors examined central AGN activity in a limited sample of 7 fairly massive (stellar mass $\gtrsim 3.9\times10^{10}$ M$_{\odot}$) cluster galaxies experiencing severe RPS effects, often referred to as `jellyfish galaxies'\footnote{These were identified from GAs Stripping Phenomena in galaxies with MUSE (GASP; \citealt{poggianti_gasp}) which is an integral field spectroscopic survey with MUSE (Multi Unit Spectroscopic Explorer) at the Very Large Telescope, focused on studying gas removal processes in galaxies. The GASP sample comprises 114 disk galaxies in the local Universe covering redshifts from 0.04 to 0.07.} due to their distinctive appearance of material extending beyond the galactic disk in tentacle-like structures spanning several kiloparsecs. Utilizing an optical emission line diagnostic approach, they found that 6 out of these 7 galaxies harbored central AGN, with 1 of these being consistent with a LINER, which may not necessarily originate from accretion onto SMBH (e.g., \citealt{belfiore2016}). Additionally, the authors reported the \textit{Chandra} broadband (0.3--8.0 keV) X-ray luminosities of the identified AGN, with most falling within the range $1-3\times 10^{41}$ erg s$^{-1}$, not indicative of unambiguous X-ray AGN activity.
 
\citet{peluso2022} expanded the GASP cluster RPS sample to include 51 galaxies (GASP-RPS), identifying 5 galaxies hosting an optical AGN (with 4 out of 5 classified as LINERs) in addition to those previously identified by \citet{poggianti2017_nature}. All of these have X-ray luminosities below $2.4\times10^{41}$ erg s$^{-1}$. Overall, the combined analysis of \citet{poggianti2017_nature} and \citet{peluso2022} suggests that only 6 out of 51 ($\sim12\%$; see Table 1 of \citet{peluso2022}) cluster RPS galaxies host an AGN (consistent with Seyfert 2 type) when excluding LINERs. Moreover, the presence of AGN or LINER in the GASP-RPS sample is limited to the most massive galaxies with stellar masses greater than $10^{10.5} \text{M}_{\odot}$ (see Fig. 2 of \citet{peluso2022}), while the sample itself covers a wide mass range of $10^{8.7} - 10^{11.5} \text{M}_{\odot}$, similar to our RPS sample. This highlights the strong dependence of presence of AGN on galaxy stellar mass, as pointed out by various authors (e.g., \citealt{kauffmann2003,decarli2007,delpino2017,yang2018,cattorini2023}). Additionally, \citet{peluso2022} compiled a heterogeneous sample of 82 cluster RPS galaxies (possibly also undergoing tidal interactions) from the literature (LIT-RPS). Among these galaxies, only 18$\%$ (15 out of 82) were identified as optical AGN, excluding LINERs (see Table 4 of \citet{peluso2022}). We note that excluding LINERs significantly reduces the proportion of galaxies hosting AGN in the GASP-RPS and LIT-RPS samples, bringing it in line with the control sample utilized by \citet{peluso2022}. \citet{boselli2022} have also emphasized the impact of including LINERs, leading to a rise in the AGN fraction in their RPS galaxy sample. Analyzing 70 nearby cluster galaxies, their study shows that 3$\%$ (2 out of 70) of late-type systems host AGN while undergoing RPS. This proportion increases to 14$\%$ (10 out of 70) when LINERs are taken into account.

The lack of correlation between AGN activity and RPS phenomenon in cluster galaxies is also corroborated by \citet{roman-oliveira2019}, who investigated a sample of 70 jellyfish galaxies within the A901/2 cluster system. Their findings revealed that only 5 galaxies ($\sim7\%$) hosted AGN. Interestingly, AGN activity was absent even among the most prominent jellyfish galaxies. Additionally, the limited number of galaxies hosting AGN were not situated in the region of the phase-space diagram where RPS is most prominent. \citet{roman-oliveira2019} also conducted a reanalysis of data from 42 GASP galaxies, revealing that the majority of these galaxies exhibit ongoing star formation unrelated to nuclear activity. A negative relationship between RPS and the level of AGN activity is also supported by \citet{delpino2023}, who examined the incidence of ionized outflows in optical AGN, commonly associated with nuclear activity, and its relation to the surrounding environment. Their study concluded that the fraction of outflows decreases at distances closer to the central galaxy of a group or cluster. Using LOFAR Two-metre Sky Survey, \citet{roberts-parker2021_rps} identified a sample of 95 jellyfish galaxies in the radio band. Their study also found no evidence for a prominent population of AGN among the jellyfish galaxies, which were classified using optical line diagnostics and mid-infrared photometry. Using the optical emission line approach, only 2$\%$ of the jellyfish galaxies were classified as Seyfert-type and this fraction increased to 7$\%$ when LINERs were taken into account. With the mid-IR photometry approach, none of the jellyfish galaxies was classified as an AGN. Similar findings were reported by \citet{durret2021} who identified only 2 out of 81 jellyfish galaxies in their sample as hosting an AGN. Likewise, \citet{kolku2022} studied a sample of jellyfish galaxies within groups and found that only 2 out of 13 galaxies with stellar masses ranging from $10^{9.5} \text{ to } 10^{10.9} \text{ M}_{\odot}$ hosted an AGN.

\subsection{Possible reasons for discrepancies in findings across different studies}
Differences in results from various studies may arise due to:\\\\
1. \textit{Sample Selection Bias}: Different studies use different criteria for selecting galaxy samples, leading to biases that influence the results. For instance, variations in the selection of RPS and control groups, inclusion/exclusion criteria, matching criteria, or sample size limited by the wavelength coverage can impact the findings.\\\\
2. \textit{Diagnostic Techniques}: Different diagnostic criteria may yield different AGN classifications and affect the inferred AGN population, hence influencing the results. Variations in the methods used to identify and classify AGN, such as emission line ratios, X-ray properties, or infrared colors, can lead to discrepancies in the observed AGN properties and their prevalence in galaxy samples. For example, \citet{cattorini2023} highlight significant differences in AGN classification between the WHAN and BPT optical emission line diagnostic diagrams. Denoting LINERs as AGN can also impact results, as discussed in section \ref{ref:subsec_compare_previous_studies}.\\\\
3. \textit{Stage of RPS}: It is possible that the cluster galaxies undergoing RPS in different samples may be at various stages of the RPS process, which could lead to observed differences in AGN activity. For example, a majority of the galaxies in the sample may have already passed through an AGN phase, while others may not have yet reached a stage where significant AGN activity is expected. Understanding at what stage of the RPS process the gas gets funneled towards the center (if such funneling happens) could help clarify differences in AGN activity among different samples. The GASP study of \citet{peluso2022} shows that galaxies with extreme tails longer than the stellar disk diameter ($JStage = 2$) tend to host AGN more frequently. It would be interesting to test this from an X-ray standpoint. Studies with larger X-ray samples, taking into consideration the RPS stages using complementary wavelengths, could provide further insight. Furthermore, larger samples would help disentangle the effects of stellar mass and RPS on AGN activity, a challenge that is currently limited by the sample size in this study. This difficulty has also been a common issue in similar studies conducted from an optical perspective (see, for e.g., \citet{peluso2022}).

\subsection{Comparison with previous studies on AGN fraction}
In our analysis of the cluster RPS and control galaxy samples, we find no significant difference in the X-ray luminosity of central point sources. This result aligns with similar studies in X-rays, which suggest that when moderate to low luminosity X-ray point sources ($<10^{42}$ erg s$^{-1}$) are included in AGN studies, the AGN fraction appears comparable between rich galaxy clusters and the field (e.g., \citealt{haggard2010}). A vast number of studies suggest that the prevalence of AGN in galaxies rises with increasing stellar mass, indicating a higher likelihood of AGN presence in more massive and brighter galaxies. However, definitive results have not been established for similar studies of AGN fraction with galaxy color. In general, X-ray studies encompassing galaxies in all environments suggest that bluer galaxies tend to have a higher AGN fraction (e.g., \citealt{silverman2009,aird2012}), while some studies (e.g., \citealt{xue2010}) do not find the AGN fraction to depend on host galaxy color. For field galaxies, the AGN fraction appears higher among bluer galaxies (e.g., \citealt{haggard2010}). Specifically, a study of AGN fraction with galaxy color in galaxy clusters may help shed more light on the effect of cluster environment on AGN activity.

\subsection{RPS-induced gas compression}
Recent observations of ram pressure stripped galaxies in clusters reveal enhanced star formation in the leading edge of these galaxies as a result of RPS-induced gas compression. However, there is no evidence of X-ray AGN signatures in these galaxies, as seen in examples such as NGC 4921 \citep{cramer2021_N4921} and IC 3949 \citep{2022_IC3949} in the Coma cluster. The absence of AGN activity may be attributed to several factors. (1) The gas compression-induced star formation may consume a large portion of the available gas, leaving insufficient material for accretion onto the central black hole to power AGN activity. (2) The timescales over which gas compression-induced star formation and AGN activity operate may not align. While star formation can proceed continuously over longer periods, AGN activity may occur in shorter, intermittent bursts. (3) While RPS-induced gas compression may primarily trigger star formation, factors other than RPS, such as galaxy mergers or interactions may be more effective at triggering AGN activity. On the simulation front, while a few recent simulation studies suggest that RPS can funnel gas towards the galaxy center and result in AGN activity, these rely on certain subgrid assumptions to account for processes that occur at scales smaller than the simulation's resolution grid. These assumptions could influence the accuracy of the simulation results (for details on caveats in simulation studies, see e.g., \citealt{ricarte2020, akerman2023}). Future observational studies with advanced telescopes, as well as high-resolution simulations incorporating more detailed physics, are expected to shed further light on the complex interplay between RPS, AGN activity, and star formation in galaxy clusters.

\subsection{Potential biases due to AGN obscuration}
A significant population of AGN ($\sim 70\%$) in the local Universe suffers from obscuration caused by dust and gas surrounding the central supermassive black hole, playing a significant role in shaping how we detect and interpret AGN activity. Based on the observed 2--10 keV count rates of the detected sources in this work, all sources (except GMP 2374 and ESO 137-002) would need to be Compton-thick, with an intrinsic column density $\gtrsim 10^{24} \text{ cm}^{-2}$, if we require their 2--10 keV nuclear X-ray luminosities to be at least $10^{42}$ erg s$^{-1}$.

If this is the case, we may be underestimating the AGN activity, particularly for the heavily obscured or Compton-thick sources. However, it is important to note that while $\sim 60-80\%$ of AGN in the local Universe are obscured, only $\sim 25-30\%$ are expected to be Compton-thick (e.g., see \citealt{malizia2009,ricci2015,akylas2016}). Given these considerations, readers should
be aware of the unexplored obscured AGN population in our sample, which is also the case for other X-ray AGN studies with {\em Chandra} and {\em XMM} (e.g., \citealt{arnold2009, haines2012, semyeong2014, koulouridis2018}).

This limitation highlights the necessity of employing multiwavelength approaches to mitigate the incompleteness in AGN samples, as no single wavelength can yield a complete AGN inventory (e.g., see \citealt{yan2011}). By integrating data across various wavelengths, we can obtain a more comprehensive understanding of AGN activity, particularly for those obscured by dense environments.
\section{Summary}
\label{sec:summary}
This study investigates the connection between AGN activity and RPS in typically late-type cluster galaxies from an X-ray perspective. We select galaxies undergoing RPS (RPS sample) from four nearby galaxy clusters -- A1656, A1367, A426, and A3627 -- and analyze signs of X-ray AGN activity within them using \textit{Chandra} data. We conduct a comparative study of X-ray AGN activity in a control sample consisting of galaxies not belonging to clusters or groups (non-RPS), matched in color, absolute magnitude, redshift, and the depth of \textit{Chandra} data with those in the RPS sample. 

Our findings suggest that most galaxies undergoing RPS have X-ray luminosities in their central point sources below $10^{41}$ erg s$^{-1}$. The most luminous X-ray AGN in the RPS sample is found within ESO 137-002, with a 2--10 keV X-ray luminosity of $\sim9\times10^{41}$ erg s$^{-1}$. We find no evidence of enhanced X-ray AGN activity in the RPS galaxies, as suggested by the similarity in X-ray luminosities observed in the central point sources of galaxies in both the RPS and control samples. This result aligns with previous studies examining X-ray AGN fraction below $10^{42}$ erg s$^{-1}$. In summary, our analysis reveals no evidence for widespread X-ray AGN activity in our cluster RPS sample, suggesting no enhancement of X-ray AGN activity from cluster environment processes such as RPS.

We note that although the sample size utilized in this study may be limited, it represents the first systematic investigation into the relationship between AGN activity and RPS from an X-ray standpoint. While we do not find broad evidence of X-ray AGN signatures in our RPS galaxy sample, conducting more X-ray studies with larger samples in the future could provide deeper insights into this connection.

While it would be interesting to put our results in the context of the general galaxy population, it is important to note that the definition of AGN fraction is often subjective to the specific choices of different studies such as observational depth, survey selection functions, and methods of AGN identification (e.g., optical versus X-ray selection). These choices can result in significant discrepancies between optically-selected and X-ray selected AGN. Moreover, a significant fraction of galaxies in our sample have stellar mass less than $10^{9.5} M_{\odot}$. The definition of AGN is murkier in these systems due to the contamination from star formation activities. In such systems, common optical AGN selection criteria such as the BPT diagnostics and the X-ray selection criteria such as the $10^{42}$ erg s$^{-1}$ threshold may not always apply (see \citealt{reines2013, birchall2020}, and reference therein for examples). As this is an area of active research, in this work, we focus on comparing the typical X-ray AGN diagnostics between our RPS sample and the carefully selected control sample to identify enhanced nuclear activity in the RPS sample. Further, more detailed observations would be required to definitively establish the AGN nature of individual targets, which is beyond the scope of this study.

\textit{Prospects}: The scientific community remains uncertain about whether RPS truly affects AGN activity, but larger X-ray samples could provide more clarity on this matter. For example, there may be differences between optical and X-ray AGN due to RPS, as mentioned earlier. Even with low accretion rates, optical AGN activity may still occur, unlike X-ray AGN. This raises questions about how much gas RPS funnels and facilitates accretion, if at all. It's possible that RPS pushes gas towards the center without resulting in high accretion rates. Further exploration in larger samples could offer more insights.

\section*{Acknowledgements}
We thank the referee for their feedback and suggestions, which have helped improve this manuscript. Support for this work was provided by the National Aeronautics and Space Administration through {\em Chandra} Award Number AR2-23013X, GO3-24077X and GO2-23082X issued by the {\em Chandra} X-ray Center, which is operated by the Smithsonian Astrophysical Observatory for and on behalf of the National Aeronautics Space Administration under contract NAS8-03060. We also acknowledge support from NSF grant 2407821.

\section*{Data Availability}
This paper employs a list of archival \textit{Chandra} datasets, obtained by the \textit{Chandra} X-ray Observatory, contained in~\dataset[DOI: https://doi.org/10.25574/cdc.295]{https://doi.org/10.25574/cdc.295}. All X-ray data are also publicly accessible via the Chandra Data Archive: \url{ https://cda.harvard.edu/chaser/}. We have used photometric data from data release 18 of the Sloan Digital Sky Survey-V. Funding for the Sloan Digital Sky Survey-V has been provided by the Alfred P. Sloan Foundation, the Heising-Simons Foundation, the National Science Foundation, and the Participating Institutions. SDSS acknowledges support and resources from the Center for High-Performance Computing at the University of Utah. SDSS telescopes are located at Apache Point Observatory, funded by the Astrophysical Research Consortium and operated by New Mexico State University, and at Las Campanas Observatory, operated by the Carnegie Institution for Science. The SDSS web site is \url{www.sdss.org}. This research has made use of SAOImageDS9, developed by Smithsonian Astrophysical Observatory, the HEASoft FTOOLS (\url{http://heasarc.gsfc.nasa.gov/ftools}), and the NASA/IPAC Extragalactic Database (NED), which is funded by the National Aeronautics and Space Administration and operated by the California Institute of Technology. The use of TOPCAT (\url{https://www.star.bristol.ac.uk/mbt/topcat/}) has been very useful for this study. Additionally, this study has utilized \textit{XMM-Newton} observations of A1656, A426, A1367 and A3627, which can be accessed at \url{http://nxsa.esac.esa.int/nxsa-web/#home}.

\bibliography{RPS-AGN.bib}

\begin{thebibliography}{}
\expandafter\ifx\csname natexlab\endcsname\relax\def\natexlab#1{#1}\fi
\providecommand{\url}[1]{\href{#1}{#1}}
\providecommand{\dodoi}[1]{doi:~\href{http://doi.org/#1}{\nolinkurl{#1}}}
\providecommand{\doeprint}[1]{\href{http://ascl.net/#1}{\nolinkurl{http://ascl.net/#1}}}
\providecommand{\doarXiv}[1]{\href{https://arxiv.org/abs/#1}{\nolinkurl{https://arxiv.org/abs/#1}}}

\bibitem[{{Aguerri} {et~al.}(2020){Aguerri}, {Girardi}, {Agulli}, {Negri},
  {Dalla Vecchia}, \& {Dom{\'\i}nguez Palmero}}]{aguerri2020}
{Aguerri}, J.~A.~L., {Girardi}, M., {Agulli}, I., {et~al.} 2020, \mnras, 494,
  1681, \dodoi{10.1093/mnras/staa800}

\bibitem[{{Aird} {et~al.}(2012){Aird}, {Coil}, {Moustakas}, {Blanton},
  {Burles}, {Cool}, {Eisenstein}, {Smith}, {Wong}, \& {Zhu}}]{aird2012}
{Aird}, J., {Coil}, A.~L., {Moustakas}, J., {et~al.} 2012, \apj, 746, 90,
  \dodoi{10.1088/0004-637X/746/1/90}

\bibitem[{{Akerman} {et~al.}(2023){Akerman}, {Tonnesen}, {Poggianti}, {Smith},
  \& {Marasco}}]{akerman2023}
{Akerman}, N., {Tonnesen}, S., {Poggianti}, B.~M., {Smith}, R., \& {Marasco},
  A. 2023, \apj, 948, 18, \dodoi{10.3847/1538-4357/acbf4d}

\bibitem[{Akylas {et~al.}(2016)Akylas, Georgantopoulos, Ranalli, Gkiokas,
  Corral, \& Lanzuisi}]{akylas2016}
Akylas, A., Georgantopoulos, I., Ranalli, P., {et~al.} 2016, \aap, 594, A73,
  \dodoi{10.1051/0004-6361/201628711}

\bibitem[{{Alexander} {et~al.}(2005){Alexander}, {Bauer}, {Chapman}, {Smail},
  {Blain}, {Brandt}, \& {Ivison}}]{alexander2005}
{Alexander}, D.~M., {Bauer}, F.~E., {Chapman}, S.~C., {et~al.} 2005, \apj, 632,
  736, \dodoi{10.1086/444342}

\bibitem[{{Alonso} {et~al.}(2007){Alonso}, {Lambas}, {Tissera}, \&
  {Coldwell}}]{alonso2007}
{Alonso}, M.~S., {Lambas}, D.~G., {Tissera}, P., \& {Coldwell}, G. 2007,
  \mnras, 375, 1017, \dodoi{10.1111/j.1365-2966.2007.11367.x}

\bibitem[{{Arnold} {et~al.}(2009){Arnold}, {Martini}, {Mulchaey}, {Berti}, \&
  {Jeltema}}]{arnold2009}
{Arnold}, T.~J., {Martini}, P., {Mulchaey}, J.~S., {Berti}, A., \& {Jeltema},
  T.~E. 2009, \apj, 707, 1691, \dodoi{10.1088/0004-637X/707/2/1691}

\bibitem[{{Bauer} {et~al.}(2004){Bauer}, {Alexander}, {Brandt}, {Schneider},
  {Treister}, {Hornschemeier}, \& {Garmire}}]{bauer2004}
{Bauer}, F.~E., {Alexander}, D.~M., {Brandt}, W.~N., {et~al.} 2004, \aj, 128,
  2048, \dodoi{10.1086/424859}

\bibitem[{{Belfiore} {et~al.}(2016){Belfiore}, {Maiolino}, {Maraston},
  {Emsellem}, {Bershady}, {Masters}, {Yan}, {Bizyaev}, {Boquien}, {Brownstein},
  {Bundy}, {Drory}, {Heckman}, {Law}, {Roman-Lopes}, {Pan}, {Stanghellini},
  {Thomas}, {Weijmans}, \& {Westfall}}]{belfiore2016}
{Belfiore}, F., {Maiolino}, R., {Maraston}, C., {et~al.} 2016, \mnras, 461,
  3111, \dodoi{10.1093/mnras/stw1234}

\bibitem[{{Best} {et~al.}(2007){Best}, {von der Linden}, {Kauffmann},
  {Heckman}, \& {Kaiser}}]{best2007}
{Best}, P.~N., {von der Linden}, A., {Kauffmann}, G., {Heckman}, T.~M., \&
  {Kaiser}, C.~R. 2007, \mnras, 379, 894,
  \dodoi{10.1111/j.1365-2966.2007.11937.x}

\bibitem[{Birchall {et~al.}(2020)Birchall, Watson, \& Aird}]{birchall2020}
Birchall, K.~L., Watson, M.~G., \& Aird, J. 2020, Monthly Notices of the Royal
  Astronomical Society, 492, 2268, \dodoi{10.1093/mnras/staa040}

\bibitem[{{Boselli} {et~al.}(2022){Boselli}, {Fossati}, \& {Sun}}]{boselli2022}
{Boselli}, A., {Fossati}, M., \& {Sun}, M. 2022, \aapr, 30, 3,
  \dodoi{10.1007/s00159-022-00140-3}

\bibitem[{{Brandt} \& {Alexander}(2015)}]{brandt2015}
{Brandt}, W.~N., \& {Alexander}, D.~M. 2015, \aapr, 23, 1,
  \dodoi{10.1007/s00159-014-0081-z}

\bibitem[{{Cattorini} {et~al.}(2023){Cattorini}, {Gavazzi}, {Boselli}, \&
  {Fossati}}]{cattorini2023}
{Cattorini}, F., {Gavazzi}, G., {Boselli}, A., \& {Fossati}, M. 2023, \aap,
  671, A118, \dodoi{10.1051/0004-6361/202244738}

\bibitem[{{Coldwell} \& {Lambas}(2006)}]{coldwell2006}
{Coldwell}, G.~V., \& {Lambas}, D.~G. 2006, \mnras, 371, 786,
  \dodoi{10.1111/j.1365-2966.2006.10712.x}

\bibitem[{{Cramer} {et~al.}(2021){Cramer}, {Kenney}, {Tonnesen}, {Smith},
  {Wong}, {J{\'a}chym}, {Cort{\'e}s}, {Cort{\'e}s}, \& {Wu}}]{cramer2021_N4921}
{Cramer}, W.~J., {Kenney}, J.~D.~P., {Tonnesen}, S., {et~al.} 2021, \apj, 921,
  22, \dodoi{10.3847/1538-4357/ac1793}

\bibitem[{{Cresci} {et~al.}(2015){Cresci}, {Marconi}, {Zibetti}, {Risaliti},
  {Carniani}, {Mannucci}, {Gallazzi}, {Maiolino}, {Balmaverde}, {Brusa},
  {Capetti}, {Cicone}, {Feruglio}, {Bland-Hawthorn}, {Nagao}, {Oliva},
  {Salvato}, {Sani}, {Tozzi}, {Urrutia}, \& {Venturi}}]{cresci2015}
{Cresci}, G., {Marconi}, A., {Zibetti}, S., {et~al.} 2015, \aap, 582, A63,
  \dodoi{10.1051/0004-6361/201526581}

\bibitem[{{Decarli} {et~al.}(2007){Decarli}, {Gavazzi}, {Arosio}, {Cortese},
  {Boselli}, {Bonfanti}, \& {Colpi}}]{decarli2007}
{Decarli}, R., {Gavazzi}, G., {Arosio}, I., {et~al.} 2007, \mnras, 381, 136,
  \dodoi{10.1111/j.1365-2966.2007.12208.x}

\bibitem[{{Donahue} \& {Voit}(2022)}]{DV22}
{Donahue}, M., \& {Voit}, G.~M. 2022, \physrep, 973, 1,
  \dodoi{10.1016/j.physrep.2022.04.005}

\bibitem[{{Dressler} {et~al.}(1985){Dressler}, {Thompson}, \&
  {Shectman}}]{dressler1985}
{Dressler}, A., {Thompson}, I.~B., \& {Shectman}, S.~A. 1985, \apj, 288, 481,
  \dodoi{10.1086/162813}

\bibitem[{{Durret} {et~al.}(2021){Durret}, {Chiche}, {Lobo}, \&
  {Jauzac}}]{durret2021}
{Durret}, F., {Chiche}, S., {Lobo}, C., \& {Jauzac}, M. 2021, \aap, 648, A63,
  \dodoi{10.1051/0004-6361/202039770}

\bibitem[{{Ehlert} {et~al.}(2014){Ehlert}, {von der Linden}, {Allen}, {Brandt},
  {Xue}, {Luo}, {Mantz}, {Morris}, {Applegate}, \& {Kelly}}]{ehlert2014}
{Ehlert}, S., {von der Linden}, A., {Allen}, S.~W., {et~al.} 2014, \mnras, 437,
  1942, \dodoi{10.1093/mnras/stt2025}

\bibitem[{{Ehlert} {et~al.}(2015){Ehlert}, {Allen}, {Brandt}, {Canning}, {Luo},
  {Mantz}, {Morris}, {von der Linden}, \& {Xue}}]{ehlert2015}
{Ehlert}, S., {Allen}, S.~W., {Brandt}, W.~N., {et~al.} 2015, \mnras, 446,
  2709, \dodoi{10.1093/mnras/stu2091}

\bibitem[{{Ellison} {et~al.}(2011){Ellison}, {Patton}, {Mendel}, \&
  {Scudder}}]{ellison2011}
{Ellison}, S.~L., {Patton}, D.~R., {Mendel}, J.~T., \& {Scudder}, J.~M. 2011,
  \mnras, 418, 2043, \dodoi{10.1111/j.1365-2966.2011.19624.x}

\bibitem[{{Fabian}(2012)}]{fabian2012}
{Fabian}, A.~C. 2012, \araa, 50, 455,
  \dodoi{10.1146/annurev-astro-081811-125521}

\bibitem[{{Farber} {et~al.}(2022){Farber}, {Ruszkowski}, {Tonnesen}, \&
  {Holguin}}]{farber2022}
{Farber}, R.~J., {Ruszkowski}, M., {Tonnesen}, S., \& {Holguin}, F. 2022,
  \mnras, 512, 5927, \dodoi{10.1093/mnras/stac794}

\bibitem[{{Fossati} {et~al.}(2012){Fossati}, {Gavazzi}, {Boselli}, \&
  {Fumagalli}}]{fossati2012_rps}
{Fossati}, M., {Gavazzi}, G., {Boselli}, A., \& {Fumagalli}, M. 2012, \aap,
  544, A128, \dodoi{10.1051/0004-6361/201219933}

\bibitem[{{Fujita} {et~al.}(2023{\natexlab{a}}){Fujita}, {Izumi}, {Kawakatu},
  {Nagai}, {Hirasawa}, \& {Ikeda}}]{fujita2023b}
{Fujita}, Y., {Izumi}, T., {Kawakatu}, N., {et~al.} 2023{\natexlab{a}}, \pasj,
  75, 925, \dodoi{10.1093/pasj/psad050}

\bibitem[{{Fujita} {et~al.}(2023{\natexlab{b}}){Fujita}, {Izumi}, {Nagai},
  {Kawakatu}, \& {Kawanaka}}]{fujita2023a}
{Fujita}, Y., {Izumi}, T., {Nagai}, H., {Kawakatu}, N., \& {Kawanaka}, N.
  2023{\natexlab{b}}, arXiv e-prints, arXiv:2310.03794,
  \dodoi{10.48550/arXiv.2310.03794}

\bibitem[{{Gao} {et~al.}(2020){Gao}, {Wang}, {Pearson}, {Gordon}, {Holwerda},
  {Hopkins}, {Brown}, {Bland-Hawthorn}, \& {Owers}}]{gao2020}
{Gao}, F., {Wang}, L., {Pearson}, W.~J., {et~al.} 2020, \aap, 637, A94,
  \dodoi{10.1051/0004-6361/201937178}

\bibitem[{{Gavazzi} {et~al.}(2018){Gavazzi}, {Consolandi}, {Gutierrez},
  {Boselli}, \& {Yoshida}}]{gavazzi2018_rps}
{Gavazzi}, G., {Consolandi}, G., {Gutierrez}, M.~L., {Boselli}, A., \&
  {Yoshida}, M. 2018, \aap, 618, A130, \dodoi{10.1051/0004-6361/201833427}

\bibitem[{{Gilmour} {et~al.}(2007){Gilmour}, {Gray}, {Almaini}, {Best}, {Wolf},
  {Meisenheimer}, {Papovich}, \& {Bell}}]{gilmour2007}
{Gilmour}, R., {Gray}, M.~E., {Almaini}, O., {et~al.} 2007, \mnras, 380, 1467,
  \dodoi{10.1111/j.1365-2966.2007.12127.x}

\bibitem[{{Gisler}(1978)}]{gisler1978}
{Gisler}, G.~R. 1978, \mnras, 183, 633, \dodoi{10.1093/mnras/183.4.633}

\bibitem[{{Goulding} {et~al.}(2018){Goulding}, {Greene}, {Bezanson}, {Greco},
  {Johnson}, {Leauthaud}, {Matsuoka}, {Medezinski}, \&
  {Price-Whelan}}]{goulding2018}
{Goulding}, A.~D., {Greene}, J.~E., {Bezanson}, R., {et~al.} 2018, \pasj, 70,
  S37, \dodoi{10.1093/pasj/psx135}

\bibitem[{{Haggard} {et~al.}(2010){Haggard}, {Green}, {Anderson}, {Constantin},
  {Aldcroft}, {Kim}, \& {Barkhouse}}]{haggard2010}
{Haggard}, D., {Green}, P.~J., {Anderson}, S.~F., {et~al.} 2010, \apj, 723,
  1447, \dodoi{10.1088/0004-637X/723/2/1447}

\bibitem[{{Haines} {et~al.}(2012){Haines}, {Pereira}, {Sanderson}, {Smith},
  {Egami}, {Babul}, {Edge}, {Finoguenov}, {Moran}, \& {Okabe}}]{haines2012}
{Haines}, C.~P., {Pereira}, M.~J., {Sanderson}, A.~J.~R., {et~al.} 2012, \apj,
  754, 97, \dodoi{10.1088/0004-637X/754/2/97}

\bibitem[{{Harrison}(2017)}]{harrison2017}
{Harrison}, C.~M. 2017, Nature Astronomy, 1, 0165,
  \dodoi{10.1038/s41550-017-0165}

\bibitem[{{Heckman} \& {Best}(2014)}]{heckman2014}
{Heckman}, T.~M., \& {Best}, P.~N. 2014, \araa, 52, 589,
  \dodoi{10.1146/annurev-astro-081913-035722}

\bibitem[{{Izumi} {et~al.}(2016){Izumi}, {Kawakatu}, \& {Kohno}}]{izumi2016}
{Izumi}, T., {Kawakatu}, N., \& {Kohno}, K. 2016, \apj, 827, 81,
  \dodoi{10.3847/0004-637X/827/1/81}

\bibitem[{{Kauffmann} {et~al.}(2004){Kauffmann}, {White}, {Heckman},
  {M{\'e}nard}, {Brinchmann}, {Charlot}, {Tremonti}, \&
  {Brinkmann}}]{kauffmann2004}
{Kauffmann}, G., {White}, S. D.~M., {Heckman}, T.~M., {et~al.} 2004, \mnras,
  353, 713, \dodoi{10.1111/j.1365-2966.2004.08117.x}

\bibitem[{{Kauffmann} {et~al.}(2003){Kauffmann}, {Heckman}, {Tremonti},
  {Brinchmann}, {Charlot}, {White}, {Ridgway}, {Brinkmann}, {Fukugita}, {Hall},
  {Ivezi{\'c}}, {Richards}, \& {Schneider}}]{kauffmann2003}
{Kauffmann}, G., {Heckman}, T.~M., {Tremonti}, C., {et~al.} 2003, \mnras, 346,
  1055, \dodoi{10.1111/j.1365-2966.2003.07154.x}

\bibitem[{{Kolcu} {et~al.}(2022){Kolcu}, {Crossett}, {Bellhouse}, \&
  {McGee}}]{kolku2022}
{Kolcu}, T., {Crossett}, J.~P., {Bellhouse}, C., \& {McGee}, S. 2022, \mnras,
  515, 5877, \dodoi{10.1093/mnras/stac2177}

\bibitem[{{Kormendy} \& {Ho}(2013)}]{kormendy2013}
{Kormendy}, J., \& {Ho}, L.~C. 2013, \araa, 51, 511,
  \dodoi{10.1146/annurev-astro-082708-101811}

\bibitem[{{Koulouridis} \& {Plionis}(2010)}]{koulouridis2010}
{Koulouridis}, E., \& {Plionis}, M. 2010, \apjl, 714, L181,
  \dodoi{10.1088/2041-8205/714/2/L181}

\bibitem[{{Koulouridis} {et~al.}(2018){Koulouridis}, {Ricci}, {Giles}, {Adami},
  {Ramos-Ceja}, {Pierre}, {Plionis}, {Lidman}, {Georgantopoulos}, {Chiappetti},
  {Elyiv}, {Ettori}, {Faccioli}, {Fotopoulou}, {Gastaldello}, {Pacaud},
  {Paltani}, \& {Vignali}}]{koulouridis2018}
{Koulouridis}, E., {Ricci}, M., {Giles}, P., {et~al.} 2018, \aap, 620, A20,
  \dodoi{10.1051/0004-6361/201832974}

\bibitem[{{Laudari} {et~al.}(2022){Laudari}, {J{\'a}chym}, {Sun}, {Waldron},
  {Chatzikos}, {Kenney}, {Luo}, {Nulsen}, {Sarazin}, {Combes}, {Edge}, {Voit},
  {Donahue}, \& {Cortese}}]{laudari2022}
{Laudari}, S., {J{\'a}chym}, P., {Sun}, M., {et~al.} 2022, \mnras, 509, 3938,
  \dodoi{10.1093/mnras/stab3280}

\bibitem[{{Lopes} {et~al.}(2017){Lopes}, {Ribeiro}, \& {Rembold}}]{lopes2017}
{Lopes}, P.~A.~A., {Ribeiro}, A.~L.~B., \& {Rembold}, S.~B. 2017, \mnras, 472,
  409, \dodoi{10.1093/mnras/stx2046}

\bibitem[{{Luo} {et~al.}(2017){Luo}, {Brandt}, {Xue}, {Lehmer}, {Alexander},
  {Bauer}, {Vito}, {Yang}, {Basu-Zych}, {Comastri}, {Gilli}, {Gu},
  {Hornschemeier}, {Koekemoer}, {Liu}, {Mainieri}, {Paolillo}, {Ranalli},
  {Rosati}, {Schneider}, {Shemmer}, {Smail}, {Sun}, {Tozzi}, {Vignali}, \&
  {Wang}}]{luo2017}
{Luo}, B., {Brandt}, W.~N., {Xue}, Y.~Q., {et~al.} 2017, \apjs, 228, 2,
  \dodoi{10.3847/1538-4365/228/1/2}

\bibitem[{{Luo} {et~al.}(2023){Luo}, {Sun}, {J{\'a}chym}, {Waldron}, {Fossati},
  {Fumagalli}, {Boselli}, {Combes}, {Kenney}, {Li}, \& {Gronke}}]{luo2023}
{Luo}, R., {Sun}, M., {J{\'a}chym}, P., {et~al.} 2023, \mnras, 521, 6266,
  \dodoi{10.1093/mnras/stad1003}

\bibitem[{{Mahajan} {et~al.}(2010){Mahajan}, {Haines}, \&
  {Raychaudhury}}]{mahajan2010}
{Mahajan}, S., {Haines}, C.~P., \& {Raychaudhury}, S. 2010, \mnras, 404, 1745,
  \dodoi{10.1111/j.1365-2966.2010.16432.x}

\bibitem[{Malizia {et~al.}(2009)Malizia, Stephen, Bassani, Bird, Panessa, \&
  Ubertini}]{malizia2009}
Malizia, A., Stephen, J.~B., Bassani, L., {et~al.} 2009, Monthly Notices of the
  Royal Astronomical Society, 399, 944,
  \dodoi{10.1111/j.1365-2966.2009.15330.x}

\bibitem[{{Marshall} {et~al.}(2018){Marshall}, {Shabala}, {Krause}, {Pimbblet},
  {Croton}, \& {Owers}}]{marshall2018}
{Marshall}, M.~A., {Shabala}, S.~S., {Krause}, M. G.~H., {et~al.} 2018, \mnras,
  474, 3615, \dodoi{10.1093/mnras/stx2996}

\bibitem[{{Martini}(2009)}]{martini2009}
{Martini}, P. 2009, in American Institute of Physics Conference Series, Vol.
  1201, The Monster's Fiery Breath: Feedback in Galaxies, Groups, and Clusters,
  ed. S.~{Heinz} \& E.~{Wilcots}, 115--118, \dodoi{10.1063/1.3293012}

\bibitem[{{McNamara} \& {Nulsen}(2007)}]{mcnamara2007}
{McNamara}, B.~R., \& {Nulsen}, P.~E.~J. 2007, \araa, 45, 117,
  \dodoi{10.1146/annurev.astro.45.051806.110625}

\bibitem[{{Miller}(2007)}]{miller2007}
{Miller}, J.~M. 2007, \araa, 45, 441,
  \dodoi{10.1146/annurev.astro.45.051806.110555}

\bibitem[{{Mishra} \& {Dai}(2020)}]{mishra2020}
{Mishra}, H.~D., \& {Dai}, X. 2020, \aj, 159, 69,
  \dodoi{10.3847/1538-3881/ab6225}

\bibitem[{{NASA/IPAC Extragalactic Database (NED)}(2019)}]{ned}
{NASA/IPAC Extragalactic Database (NED)}. 2019, NASA/IPAC Extragalactic
  Database (NED),  IPAC, \dodoi{10.26132/NED1}

\bibitem[{{Oh} {et~al.}(2014){Oh}, {Mulchaey}, {Woo}, {Finoguenov}, {Tanaka},
  {Cooper}, {Ziparo}, {Bauer}, \& {Matsuoka}}]{semyeong2014}
{Oh}, S., {Mulchaey}, J.~S., {Woo}, J.-H., {et~al.} 2014, \apj, 790, 43,
  \dodoi{10.1088/0004-637X/790/1/43}

\bibitem[{{Padovani} {et~al.}(2017){Padovani}, {Alexander}, {Assef}, {De
  Marco}, {Giommi}, {Hickox}, {Richards}, {Smol{\v{c}}i{\'c}},
  {Hatziminaoglou}, {Mainieri}, \& {Salvato}}]{padovani2017}
{Padovani}, P., {Alexander}, D.~M., {Assef}, R.~J., {et~al.} 2017, \aapr, 25,
  2, \dodoi{10.1007/s00159-017-0102-9}

\bibitem[{{Parkash} {et~al.}(2019){Parkash}, {Brown}, {Jarrett},
  {Fraser-McKelvie}, \& {Cluver}}]{parkash2019}
{Parkash}, V., {Brown}, M. J.~I., {Jarrett}, T.~H., {Fraser-McKelvie}, A., \&
  {Cluver}, M.~E. 2019, \mnras, 485, 3169, \dodoi{10.1093/mnras/stz593}

\bibitem[{{Peluso} {et~al.}(2022){Peluso}, {Vulcani}, {Poggianti}, {Moretti},
  {Radovich}, {Smith}, {Jaff{\'e}}, {Crossett}, {Gullieuszik}, {Fritz}, \&
  {Ignesti}}]{peluso2022}
{Peluso}, G., {Vulcani}, B., {Poggianti}, B.~M., {et~al.} 2022, \apj, 927, 130,
  \dodoi{10.3847/1538-4357/ac4225}

\bibitem[{{Poggianti} {et~al.}(2017{\natexlab{a}}){Poggianti}, {Jaff{\'e}},
  {Moretti}, {Gullieuszik}, {Radovich}, {Tonnesen}, {Fritz}, {Bettoni},
  {Vulcani}, {Fasano}, {Bellhouse}, {Hau}, \&
  {Omizzolo}}]{poggianti2017_nature}
{Poggianti}, B.~M., {Jaff{\'e}}, Y.~L., {Moretti}, A., {et~al.}
  2017{\natexlab{a}}, \nat, 548, 304, \dodoi{10.1038/nature23462}

\bibitem[{{Poggianti} {et~al.}(2017{\natexlab{b}}){Poggianti}, {Moretti},
  {Gullieuszik}, {Fritz}, {Jaff{\'e}}, {Bettoni}, {Fasano}, {Bellhouse}, {Hau},
  {Vulcani}, {Biviano}, {Omizzolo}, {Paccagnella}, {D'Onofrio}, {Cava},
  {Sheen}, {Couch}, \& {Owers}}]{poggianti_gasp}
{Poggianti}, B.~M., {Moretti}, A., {Gullieuszik}, M., {et~al.}
  2017{\natexlab{b}}, \apj, 844, 48, \dodoi{10.3847/1538-4357/aa78ed}

\bibitem[{{Popesso} \& {Biviano}(2006)}]{popesso2006}
{Popesso}, P., \& {Biviano}, A. 2006, \aap, 460, L23,
  \dodoi{10.1051/0004-6361:20066269}

\bibitem[{{Reines} {et~al.}(2013){Reines}, {Greene}, \& {Geha}}]{reines2013}
{Reines}, A.~E., {Greene}, J.~E., \& {Geha}, M. 2013, \apj, 775, 116,
  \dodoi{10.1088/0004-637X/775/2/116}

\bibitem[{{Ricarte} {et~al.}(2020){Ricarte}, {Tremmel}, {Natarajan}, \&
  {Quinn}}]{ricarte2020}
{Ricarte}, A., {Tremmel}, M., {Natarajan}, P., \& {Quinn}, T. 2020, \apjl, 895,
  L8, \dodoi{10.3847/2041-8213/ab9022}

\bibitem[{Ricci {et~al.}(2015)Ricci, Ueda, Koss, Trakhtenbrot, Bauer, \&
  Gandhi}]{ricci2015}
Ricci, C., Ueda, Y., Koss, M.~J., {et~al.} 2015, The Astrophysical Journal
  Letters, 815, L13, \dodoi{10.1088/2041-8205/815/1/L13}

\bibitem[{{Roberts} \& {Parker}(2020)}]{roberts-parker2020_rps}
{Roberts}, I.~D., \& {Parker}, L.~C. 2020, \mnras, 495, 554,
  \dodoi{10.1093/mnras/staa1213}

\bibitem[{{Roberts} {et~al.}(2022{\natexlab{a}}){Roberts}, {van Weeren},
  {Timmerman}, {Botteon}, {Gendron-Marsolais}, {Ignesti}, \&
  {Rottgering}}]{roberts-parker2022_rps}
{Roberts}, I.~D., {van Weeren}, R.~J., {Timmerman}, R., {et~al.}
  2022{\natexlab{a}}, \aap, 658, A44, \dodoi{10.1051/0004-6361/202142294}

\bibitem[{{Roberts} {et~al.}(2021){Roberts}, {van Weeren}, {McGee}, {Botteon},
  {Drabent}, {Ignesti}, {Rottgering}, {Shimwell}, \&
  {Tasse}}]{roberts-parker2021_rps}
{Roberts}, I.~D., {van Weeren}, R.~J., {McGee}, S.~L., {et~al.} 2021, \aap,
  650, A111, \dodoi{10.1051/0004-6361/202140784}

\bibitem[{{Roberts} {et~al.}(2022{\natexlab{b}}){Roberts}, {Lang}, {Trotsenko},
  {Bemis}, {Ellison}, {Lin}, {Pan}, {Ignesti}, {Leslie}, \& {van
  Weeren}}]{2022_IC3949}
{Roberts}, I.~D., {Lang}, M., {Trotsenko}, D., {et~al.} 2022{\natexlab{b}},
  \apj, 941, 77, \dodoi{10.3847/1538-4357/ac9e9f}

\bibitem[{{Rodr{\'\i}guez Del Pino} {et~al.}(2023){Rodr{\'\i}guez Del Pino},
  {Arribas}, {Chies-Santos}, {Lamperti}, {Perna}, \&
  {V{\'\i}lchez}}]{delpino2023}
{Rodr{\'\i}guez Del Pino}, B., {Arribas}, S., {Chies-Santos}, A.~L., {et~al.}
  2023, \aap, 675, A41, \dodoi{10.1051/0004-6361/202346051}

\bibitem[{{Rodr{\'\i}guez del Pino} {et~al.}(2017){Rodr{\'\i}guez del Pino},
  {Arag{\'o}n-Salamanca}, {Chies-Santos}, {Weinzirl}, {Bamford}, {Gray},
  {B{\"o}hm}, {Wolf}, \& {Maltby}}]{delpino2017}
{Rodr{\'\i}guez del Pino}, B., {Arag{\'o}n-Salamanca}, A., {Chies-Santos},
  A.~L., {et~al.} 2017, \mnras, 467, 4200, \dodoi{10.1093/mnras/stx228}

\bibitem[{{Roman-Oliveira} {et~al.}(2019){Roman-Oliveira}, {Chies-Santos},
  {Rodr{\'\i}guez del Pino}, {Arag{\'o}n-Salamanca}, {Gray}, \&
  {Bamford}}]{roman-oliveira2019}
{Roman-Oliveira}, F.~V., {Chies-Santos}, A.~L., {Rodr{\'\i}guez del Pino}, B.,
  {et~al.} 2019, \mnras, 484, 892, \dodoi{10.1093/mnras/stz007}

\bibitem[{{Sabater} {et~al.}(2013){Sabater}, {Best}, \&
  {Argudo-Fern{\'a}ndez}}]{sabater2013}
{Sabater}, J., {Best}, P.~N., \& {Argudo-Fern{\'a}ndez}, M. 2013, \mnras, 430,
  638, \dodoi{10.1093/mnras/sts675}

\bibitem[{{Sabater} {et~al.}(2015){Sabater}, {Best}, \&
  {Heckman}}]{sabater2015}
{Sabater}, J., {Best}, P.~N., \& {Heckman}, T.~M. 2015, \mnras, 447, 110,
  \dodoi{10.1093/mnras/stu2429}

\bibitem[{{Satyapal} {et~al.}(2014){Satyapal}, {Ellison}, {McAlpine}, {Hickox},
  {Patton}, \& {Mendel}}]{satyapal2014}
{Satyapal}, S., {Ellison}, S.~L., {McAlpine}, W., {et~al.} 2014, \mnras, 441,
  1297, \dodoi{10.1093/mnras/stu650}

\bibitem[{{Shin} {et~al.}(2019){Shin}, {Woo}, {Chung}, {Baek}, {Cho}, {Kang},
  \& {Bae}}]{shin2019}
{Shin}, J., {Woo}, J.-H., {Chung}, A., {et~al.} 2019, \apj, 881, 147,
  \dodoi{10.3847/1538-4357/ab2e72}

\bibitem[{{Silverman} {et~al.}(2009){Silverman}, {Lamareille}, {Maier},
  {Lilly}, {Mainieri}, {Brusa}, {Cappelluti}, {Hasinger}, {Zamorani},
  {Scodeggio}, {Bolzonella}, {Contini}, {Carollo}, {Jahnke}, {Kneib}, {Le
  F{\`e}vre}, {Merloni}, {Bardelli}, {Bongiorno}, {Brunner}, {Caputi},
  {Civano}, {Comastri}, {Coppa}, {Cucciati}, {de la Torre}, {de Ravel},
  {Elvis}, {Finoguenov}, {Fiore}, {Franzetti}, {Garilli}, {Gilli}, {Iovino},
  {Kampczyk}, {Knobel}, {Kova{\v{c}}}, {Le Borgne}, {Le Brun}, {Mignoli},
  {Pello}, {Peng}, {Perez Montero}, {Ricciardelli}, {Tanaka}, {Tasca},
  {Tresse}, {Vergani}, {Vignali}, {Zucca}, {Bottini}, {Cappi}, {Cassata},
  {Fumana}, {Griffiths}, {Kartaltepe}, {Koekemoer}, {Marinoni}, {McCracken},
  {Memeo}, {Meneux}, {Oesch}, {Porciani}, \& {Salvato}}]{silverman2009}
{Silverman}, J.~D., {Lamareille}, F., {Maier}, C., {et~al.} 2009, \apj, 696,
  396, \dodoi{10.1088/0004-637X/696/1/396}

\bibitem[{{Silverman} {et~al.}(2011){Silverman}, {Kampczyk}, {Jahnke},
  {Andrae}, {Lilly}, {Elvis}, {Civano}, {Mainieri}, {Vignali}, {Zamorani},
  {Nair}, {Le F{\`e}vre}, {de Ravel}, {Bardelli}, {Bongiorno}, {Bolzonella},
  {Cappi}, {Caputi}, {Carollo}, {Contini}, {Coppa}, {Cucciati}, {de la Torre},
  {Franzetti}, {Garilli}, {Halliday}, {Hasinger}, {Iovino}, {Knobel},
  {Koekemoer}, {Kova{\v{c}}}, {Lamareille}, {Le Borgne}, {Le Brun}, {Maier},
  {Mignoli}, {Pello}, {P{\'e}rez-Montero}, {Ricciardelli}, {Peng}, {Scodeggio},
  {Tanaka}, {Tasca}, {Tresse}, {Vergani}, {Zucca}, {Brusa}, {Cappelluti},
  {Comastri}, {Finoguenov}, {Fu}, {Gilli}, {Hao}, {Ho}, \&
  {Salvato}}]{silverman2011}
{Silverman}, J.~D., {Kampczyk}, P., {Jahnke}, K., {et~al.} 2011, \apj, 743, 2,
  \dodoi{10.1088/0004-637X/743/1/2}

\bibitem[{{Smith} {et~al.}(2010){Smith}, {Lucey}, {Hammer}, {Hornschemeier},
  {Carter}, {Hudson}, {Marzke}, {Mouhcine}, {Eftekharzadeh}, {James},
  {Khosroshahi}, {Kourkchi}, \& {Karick}}]{smith2010_rps}
{Smith}, R.~J., {Lucey}, J.~R., {Hammer}, D., {et~al.} 2010, \mnras, 408, 1417,
  \dodoi{10.1111/j.1365-2966.2010.17253.x}

\bibitem[{{Smith} {et~al.}(2001){Smith}, {Brickhouse}, {Liedahl}, \&
  {Raymond}}]{smith2001}
{Smith}, R.~K., {Brickhouse}, N.~S., {Liedahl}, D.~A., \& {Raymond}, J.~C.
  2001, \apjl, 556, L91, \dodoi{10.1086/322992}

\bibitem[{{Sohn} {et~al.}(2020){Sohn}, {Geller}, {Diaferio}, \&
  {Rines}}]{sohn2020}
{Sohn}, J., {Geller}, M.~J., {Diaferio}, A., \& {Rines}, K.~J. 2020, \apj, 891,
  129, \dodoi{10.3847/1538-4357/ab6e6a}

\bibitem[{{Sun} {et~al.}(2010){Sun}, {Donahue}, {Roediger}, {Nulsen}, {Voit},
  {Sarazin}, {Forman}, \& {Jones}}]{sun2010_A3627}
{Sun}, M., {Donahue}, M., {Roediger}, E., {et~al.} 2010, \apj, 708, 946,
  \dodoi{10.1088/0004-637X/708/2/946}

\bibitem[{{Sun} {et~al.}(2007){Sun}, {Donahue}, \& {Voit}}]{sun2007}
{Sun}, M., {Donahue}, M., \& {Voit}, G.~M. 2007, \apj, 671, 190,
  \dodoi{10.1086/522690}

\bibitem[{{Tempel} {et~al.}(2014){Tempel}, {Tamm}, {Gramann}, {Tuvikene},
  {Liivam{\"a}gi}, {Suhhonenko}, {Kipper}, {Einasto}, \& {Saar}}]{tempel2014}
{Tempel}, E., {Tamm}, A., {Gramann}, M., {et~al.} 2014, \aap, 566, A1,
  \dodoi{10.1051/0004-6361/201423585}

\bibitem[{{Waldron} {et~al.}(2023){Waldron}, {Sun}, {Luo}, {Laudari},
  {Chatzikos}, {Sivanandam}, {Kenney}, {J{\'a}chym}, {Voit}, {Donahue}, \&
  {Fossati}}]{waldron2023}
{Waldron}, W., {Sun}, M., {Luo}, R., {et~al.} 2023, \mnras, 522, 173,
  \dodoi{10.1093/mnras/stad963}

\bibitem[{{Willingale} {et~al.}(2013){Willingale}, {Starling}, {Beardmore},
  {Tanvir}, \& {O'Brien}}]{willingale2013}
{Willingale}, R., {Starling}, R.~L.~C., {Beardmore}, A.~P., {Tanvir}, N.~R., \&
  {O'Brien}, P.~T. 2013, \mnras, 431, 394, \dodoi{10.1093/mnras/stt175}

\bibitem[{{Wilms} {et~al.}(2000){Wilms}, {Allen}, \& {McCray}}]{wilms2000}
{Wilms}, J., {Allen}, A., \& {McCray}, R. 2000, \apj, 542, 914,
  \dodoi{10.1086/317016}

\bibitem[{{Woudt} {et~al.}(2008){Woudt}, {Kraan-Korteweg}, {Lucey}, {Fairall},
  \& {Moore}}]{woudt2008}
{Woudt}, P.~A., {Kraan-Korteweg}, R.~C., {Lucey}, J., {Fairall}, A.~P., \&
  {Moore}, S.~A.~W. 2008, \mnras, 383, 445,
  \dodoi{10.1111/j.1365-2966.2007.12571.x}

\bibitem[{{Xue} {et~al.}(2016){Xue}, {Luo}, {Brandt}, {Alexander}, {Bauer},
  {Lehmer}, \& {Yang}}]{xue2016}
{Xue}, Y.~Q., {Luo}, B., {Brandt}, W.~N., {et~al.} 2016, \apjs, 224, 15,
  \dodoi{10.3847/0067-0049/224/2/15}

\bibitem[{{Xue} {et~al.}(2010){Xue}, {Brandt}, {Luo}, {Rafferty}, {Alexander},
  {Bauer}, {Lehmer}, {Schneider}, \& {Silverman}}]{xue2010}
{Xue}, Y.~Q., {Brandt}, W.~N., {Luo}, B., {et~al.} 2010, \apj, 720, 368,
  \dodoi{10.1088/0004-637X/720/1/368}

\bibitem[{{Xue} {et~al.}(2011){Xue}, {Luo}, {Brandt}, {Bauer}, {Lehmer},
  {Broos}, {Schneider}, {Alexander}, {Brusa}, {Comastri}, {Fabian}, {Gilli},
  {Hasinger}, {Hornschemeier}, {Koekemoer}, {Liu}, {Mainieri}, {Paolillo},
  {Rafferty}, {Rosati}, {Shemmer}, {Silverman}, {Smail}, {Tozzi}, \&
  {Vignali}}]{xue2011}
{Xue}, Y.~Q., {Luo}, B., {Brandt}, W.~N., {et~al.} 2011, \apjs, 195, 10,
  \dodoi{10.1088/0067-0049/195/1/10}

\bibitem[{{Yagi} {et~al.}(2017){Yagi}, {Yoshida}, {Gavazzi}, {Komiyama},
  {Kashikawa}, \& {Okamura}}]{yagi2017_rps}
{Yagi}, M., {Yoshida}, M., {Gavazzi}, G., {et~al.} 2017, \apj, 839, 65,
  \dodoi{10.3847/1538-4357/aa68e3}

\bibitem[{{Yagi} {et~al.}(2010){Yagi}, {Yoshida}, {Komiyama}, {Kashikawa},
  {Furusawa}, {Okamura}, {Graham}, {Miller}, {Carter}, {Mobasher}, \&
  {Jogee}}]{yagi2010_rps}
{Yagi}, M., {Yoshida}, M., {Komiyama}, Y., {et~al.} 2010, \aj, 140, 1814,
  \dodoi{10.1088/0004-6256/140/6/1814}

\bibitem[{{Yan} \& {Blanton}(2012)}]{yan2012}
{Yan}, R., \& {Blanton}, M.~R. 2012, \apj, 747, 61,
  \dodoi{10.1088/0004-637X/747/1/61}

\bibitem[{Yan {et~al.}(2011)Yan, Ho, Newman, Coil, Willmer, Laird, Georgakakis,
  Aird, Barmby, Bundy, Cooper, Davis, Faber, Fang, Griffith, Koekemoer, Koo,
  Nandra, Park, Sarajedini, Weiner, \& Willner}]{yan2011}
Yan, R., Ho, L.~C., Newman, J.~A., {et~al.} 2011, The Astrophysical Journal,
  728, 38, \dodoi{10.1088/0004-637X/728/1/38}

\bibitem[{{Yang} {et~al.}(2018){Yang}, {Brandt}, {Darvish}, {Chen}, {Vito},
  {Alexander}, {Bauer}, \& {Trump}}]{yang2018}
{Yang}, G., {Brandt}, W.~N., {Darvish}, B., {et~al.} 2018, \mnras, 480, 1022,
  \dodoi{10.1093/mnras/sty1910}

\bibitem[{{Yesuf} \& {Ho}(2020)}]{yesuf2020}
{Yesuf}, H.~M., \& {Ho}, L.~C. 2020, \apj, 901, 42,
  \dodoi{10.3847/1538-4357/aba961}

\bibitem[{{Zhang} {et~al.}(2013){Zhang}, {Sun}, {Ji}, {Sarazin}, {Lin},
  {Nulsen}, {Roediger}, {Donahue}, {Forman}, {Jones}, {Voit}, \&
  {Kong}}]{zhang2013}
{Zhang}, B., {Sun}, M., {Ji}, L., {et~al.} 2013, \apj, 777, 122,
  \dodoi{10.1088/0004-637X/777/2/122}

\end{thebibliography}
\bibliographystyle{aasjournal}

\end{document}